\documentclass[preprint,aoas]{imsart2}

\RequirePackage{amsthm,amsmath,amsfonts,amssymb}
\RequirePackage{algorithm,algorithmic}

\RequirePackage[draft=false,breaklinks=true,colorlinks,linktocpage,citecolor=blue,linkcolor=blue,urlcolor=blue]{hyperref}
\RequirePackage{breakurl}
\RequirePackage{verbatim}
\RequirePackage{color}
\RequirePackage{xcolor} 
\usepackage{url}

\RequirePackage[pdftex]{graphicx}
\RequirePackage{float}

\RequirePackage{multirow}
\RequirePackage{eqparbox}

\RequirePackage{natbib}

\usepackage{xr}
\pdfoutput=1

\usepackage{framed}

\startlocaldefs
\numberwithin{equation}{section}
\theoremstyle{plain}

\endlocaldefs

\begin{document}


\begin{frontmatter}
\title{Achieving a Hyperlocal Housing Price Index: Overcoming Data Sparsity by Bayesian Dynamical Modeling of Multiple Data Streams}
\runtitle{Modeling a Hyperlocal Housing Price Index}

\begin{aug}
\author{\fnms{You} \snm{Ren}\thanksref{wash}\ead[label=e1]{shirleyr@uw.edu}},
\author{\fnms{Emily B.} \snm{Fox}\thanksref{wash}\corref{}\ead[label=e2]{ebfox@stat.washington.edu}},
\and
\author{\fnms{Andrew} \snm{Bruce}\thanksref{wash}\ead[label=e3]{andrewb0@uw.edu}}

\runauthor{Y. Ren et al.}

\affiliation{University of Washington\thanksmark{wash}}

\address{You Ren, Emily B. Fox, and Andrew Bruce\\
Department of Statistics\\
University of Washington\\
Box 354322\\
Seattle, WA 98195\\
USA\\
\printead{e1}\\
\phantom{E-mail:\ }\printead*{e2}\\
\phantom{E-mail:\ }\printead*{e3}
}

\end{aug}

\begin{abstract}
Understanding how housing values evolve over time is important to policy makers, consumers and real estate professionals. Existing methods for constructing housing indices are computed at a coarse spatial granularity, such as metropolitan regions, which can mask or distort price dynamics apparent in local markets, such as neighborhoods and census tracts.  A challenge in moving to estimates at, for example, the census tract level is the sparsity of spatiotemporally localized house sales observations. Our work aims at addressing this challenge by leveraging observations from multiple census tracts discovered to have correlated valuation dynamics. Our proposed Bayesian nonparametric approach builds on the framework of latent factor models to enable a flexible, data-driven method for inferring the clustering of correlated census tracts.  
We explore methods for scalability and parallelizability of computations, yielding a housing valuation index at the level of census tract rather than zip code, and on a monthly basis rather than quarterly. Our analysis is provided on a large Seattle metropolitan housing dataset.
\end{abstract}

%

\end{frontmatter}

\section{Introduction}\label{sec1:introduction}
The housing market is a large part of the global economy. In the United States, roughly fifty percent of household wealth is in residential real estate, according to a Federal Reserve Study \citep{FedStudy2011}.  Between 15\% and 17\% of the U.S. gross domestic product is on housing and housing related services according to GDP statistics published by the U.S. Bureau of Economic Analysis.  Understanding how the value of housing changes over time is important to policy makers, consumers, real estate professionals and mortgage lenders.  Valuation is relatively straightforward for commoditized sectors of the economy, such as energy or non-discretionary spending.  By contrast, valuation of residential real estate is intrinsically difficult due to the individual nature of houses. Since the composition of the houses sold changes from one time period to the next, the change in the reported prices does not necessarily reflect the overall change in value.  Consequently, economists and public policy researchers have devoted considerable effort to developing a meaningful index to measure the change in housing prices over time.

The most common approach to constructing a housing price index is the repeat sales model, first proposed by \cite{RepeatSales1963}. The main idea is to use a pair of sales for the same house to model the price trend over time. Assuming the house remains in the same condition, the first sales price serves as a surrogate for the house \emph{hedonics} (house-level covariates) and the difference in the subsequent sales price captures the change in value over that intra-sales period.  This approach largely circumvents the problem caused by the change in composition of houses sold. A large body of literature extends the original repeat sales model with numerous modifications and improvements~\citep[cf.,][]{CaseShiller1987,CaseShiller1989,Gatzlaff1997,Shiller1991,Goetzmann2002}. The repeat sales model is the basis for the Case-Shiller home value index, published by Core-Logic 
and widely disseminated by the media.

One drawback of a repeat sales model is that houses with only a single sales transaction get discarded from the dataset. \cite{CaseShiller1987} report that, over a study period of 16 years, single sales make up as much as 93\%-97\% of total transactions for metropolitan areas such as Atlanta, Dallas, Chicago and San Francisco. As such, studies based on repeat sales data rely on only a fraction of all transactions and may not be a good representation of the entire house market. \cite{Englund1999} and \cite{MeeseWallace1997} detected a sampling selection bias in which the repeat sales properties are older, smaller and more modest than single-sale properties. Furthermore, small samples lead to less precise parameter estimation. To overcome this, \cite{CaseQuigley1991} propose a hybrid model that combines repeat sales with hedonic information to make use of all sales.  Recently, \cite{Brown2011} propose an autoregressive repeat sales model that utilizes all sales data without the need for hedonic information. Their approach leads to an index estimated quarterly at the zip code level. 

Existing repeat sales models, even those using all of the transactions, perform the best when fit to relatively large areas, such as metropolitan areas or cities.  Despite the large number of house sales observations in aggregate, when considering fine spatial resolutions, such as neighborhoods or census tracts, we have a large $p$ (number of regions) small $n$ (number of spatiotemporally-localized sales) problem.  For example, in our dataset described in Section~\ref{sec2:data}, most census tracts (114 out of 140) have fewer than 5 sales per month on average (see Table~\ref{tab:obsFrequency}).  The sparsity of transactions makes it challenging to obtain stable parameter estimates for small regions, and repeat sales models lack stability and predictive accuracy.  This is a significant limitation: the value of real estate is intrinsically local and coarse-scale estimates may mask or distort key phenomena.
\begin{table}[!hb]
\caption{Number of census tracts in Seattle City that have less than single digit transactions per month on average.}
\label{tab:obsFrequency}
\begin{tabular}{lrrrrr}
\hline
Average monthly sales & $<1$ & $<3$ & $<5$ & $<7$ & $<9$ \\
\hline
Number of tracts & 16 & 58 & 114 & 136 & 139 \\
Percentage of tracts &   0.11 & 0.41 & 0.81 & 0.97 & 0.99 
\\ \hline
\end{tabular}
\end{table}

An alternative, bottom-up, approach to constructing a housing price index is to compute an estimate of each individual house value and then aggregate the house-level estimates. Zillow pioneered this approach with the Zillow Home Value Index (ZHVI\textsuperscript{\textregistered}) \citep{ZHVImethodology} by taking the median of all house-level estimates  (Zestimate\textsuperscript{\textregistered}) within  a given region. The ZHVI is appealing due to its straightforward and intuitive nature. Unlike weighted repeat sales methods, the ZHVI is not impacted by the changing composition in types of homes that are sold over different periods of time. In addition, the ZHVI is stable for even very small geographic regions, such as a census tract. While the ZHVI confers certain advantages, there are limitations with the method. The approach is empirical in nature, and as such, does not directly try to model the underlying spatiotemporal dynamics of house values.  House-level estimates are based on a prediction model proprietary to Zillow that uses a variety of data from different sources.  The most important data are recent transactions.  Depending on the homogeneity of the homes in an area and the uniqueness of a particular home, a significant history of transactions may be needed for a reliable estimate.  This is a problem because the prediction model needs to adjust for the time of sale in order to account for the change in home value over time. In other words, the accuracy of the house-level prediction model, and consequently the ZHVI, is dependent on how well it captures the spatiotemporal dynamics of house values.

The main contribution of this paper is developing a model-based approach to creating housing indices on a finer spatiotemporal granularity than current methods. The indices are valuable for direct analysis and also as input into house-level models.
Our formulation is based on a dynamic model that introduces a latent process to capture the census-tract-level housing valuation index on a monthly basis (although the ideas scale to finer spatiotemporal resolutions). This latent process is informed by all individual house sales within the census tract, including detailed information of sales prices and house hedonics.  To overcome the sparseness of sales within a census tract, we inform the latent price trends based on sales in multiple census tracts discovered to have similar dynamics.  

\begin{figure}
\centering
\begin{tabular}{@{\hspace{0mm}}c@{\hspace{1mm}}c@{\hspace{1mm}}c@{\hspace{1mm}}c@{\hspace{1mm}}c@{\hspace{1mm}}c@{\hspace{1mm}}c}
\begin{tabular}{@{\hspace{0mm}}c@{\hspace{1mm}}c@{\hspace{1mm}}c@{\hspace{1mm}}c@{\hspace{1mm}}c@{\hspace{1mm}}c@{\hspace{1mm}}c}
\includegraphics[width=0.55\linewidth]{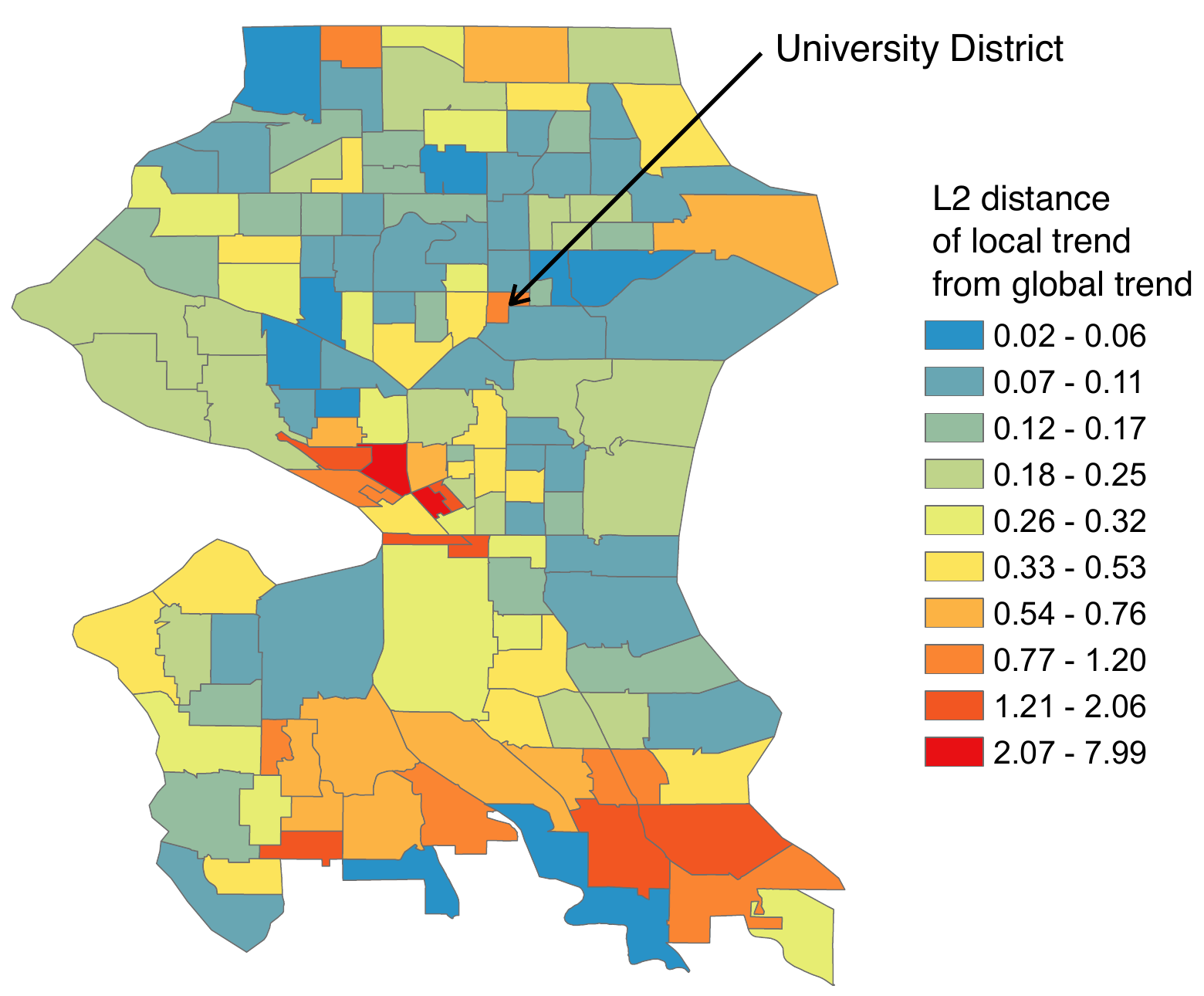}&
\includegraphics[width=0.44\linewidth]{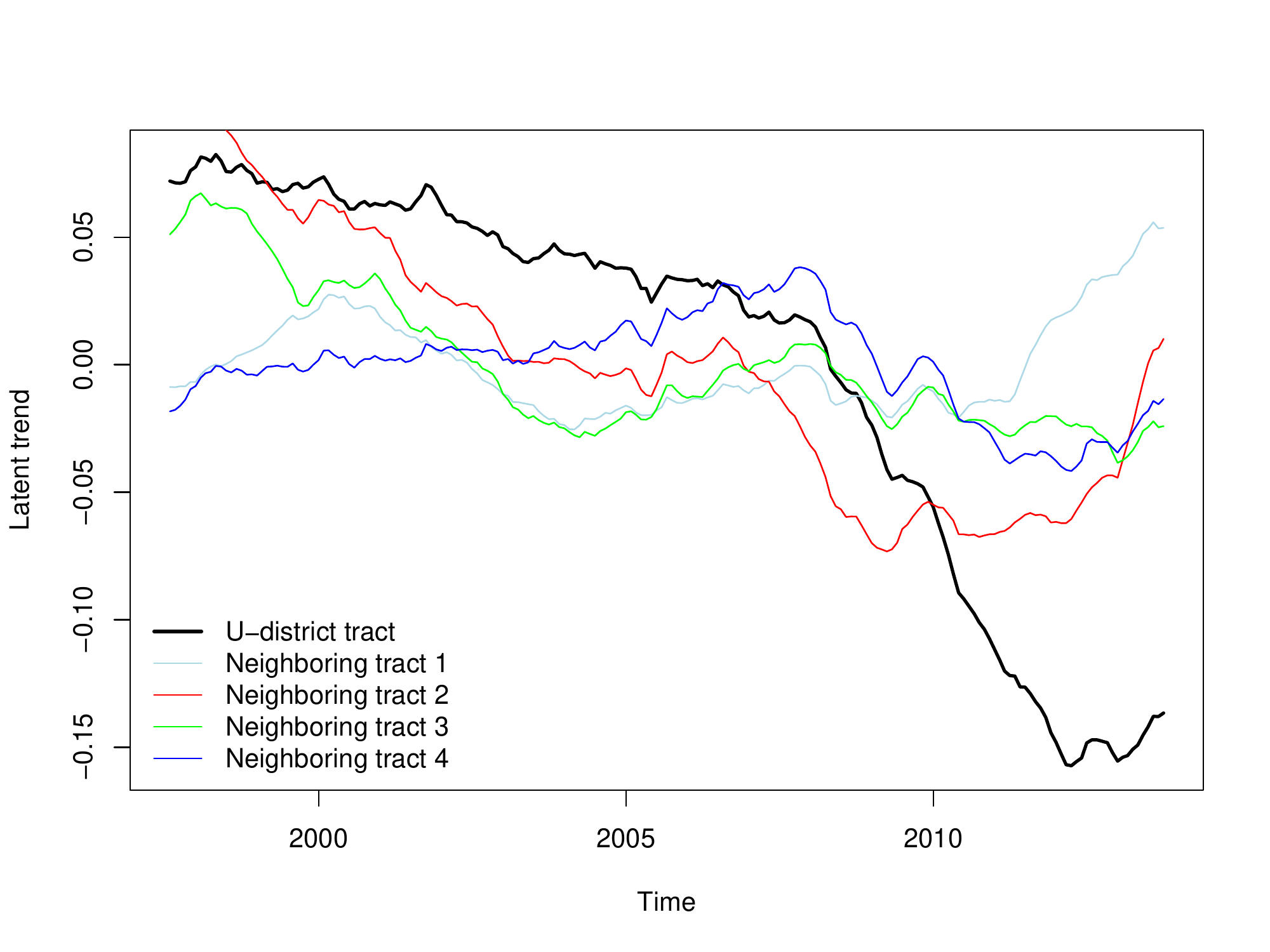}\\
(a) & (b)
\end{tabular}\\
\end{tabular}
\caption{(a) Map of inferred tract-specific latent price dynamics, where the color shows how different the local trend is from the global trend, measured in $L2$ distance over time. (b) The University District's latent price dynamics ({\em black}), which vary significantly from its neighboring census tracts ({\em other colors}). More details are in Section~\ref{sec9:realData}.}
\label{fig:Map_latentTrend}
\end{figure}

Unlike many spatiotemporal processes, modeling the correlation using Euclidean distance is not appropriate since spatially disjoint regions can be quite similar while neighboring census tracts can have significantly different value dynamics. 
For example, census tracts adjacent to waterfront, even if far apart, tend to share more in valuation dynamics than nearby census tracts that are not adjacent to waterfront. In our analysis of house sales in Seattle described in Section~\ref{sec9:realData}, we indeed find that certain census tracts vary dramatically from neighboring census tracts. Figure~\ref{fig:Map_latentTrend}(a) shows a map of deviations of each census tract's inferred local price dynamics from a global trend.  We clearly see spatially abrupt changes between neighboring regions.  One example in Figure~\ref{fig:Map_latentTrend}(a) is the University District (U-District).  Figure~\ref{fig:Map_latentTrend}(b) shows that the price trend in the U-District behaves differently compared to its neighboring census tracts.  This census tract is heavily populated by University of Washington students and has a higher crime rate than neighboring tracts. Instead of relying on an explicit spatial model, we develop a Bayesian nonparametric clustering approach to infer the relational structure of the census tracts based solely on observed house sales prices (after accounting for associated hedonics).  Within a cluster, the latent value dynamics are correlated whereas census tracts in different clusters are assumed to evolve independently.  By leveraging Bayesian nonparametrics---specifically building on the Dirichlet process---our formulation enables a flexible, data-driven method for discovering these clustered dynamics, including the number of clusters.

\begin{figure}[t!]
\centering
\includegraphics[width=1\linewidth]{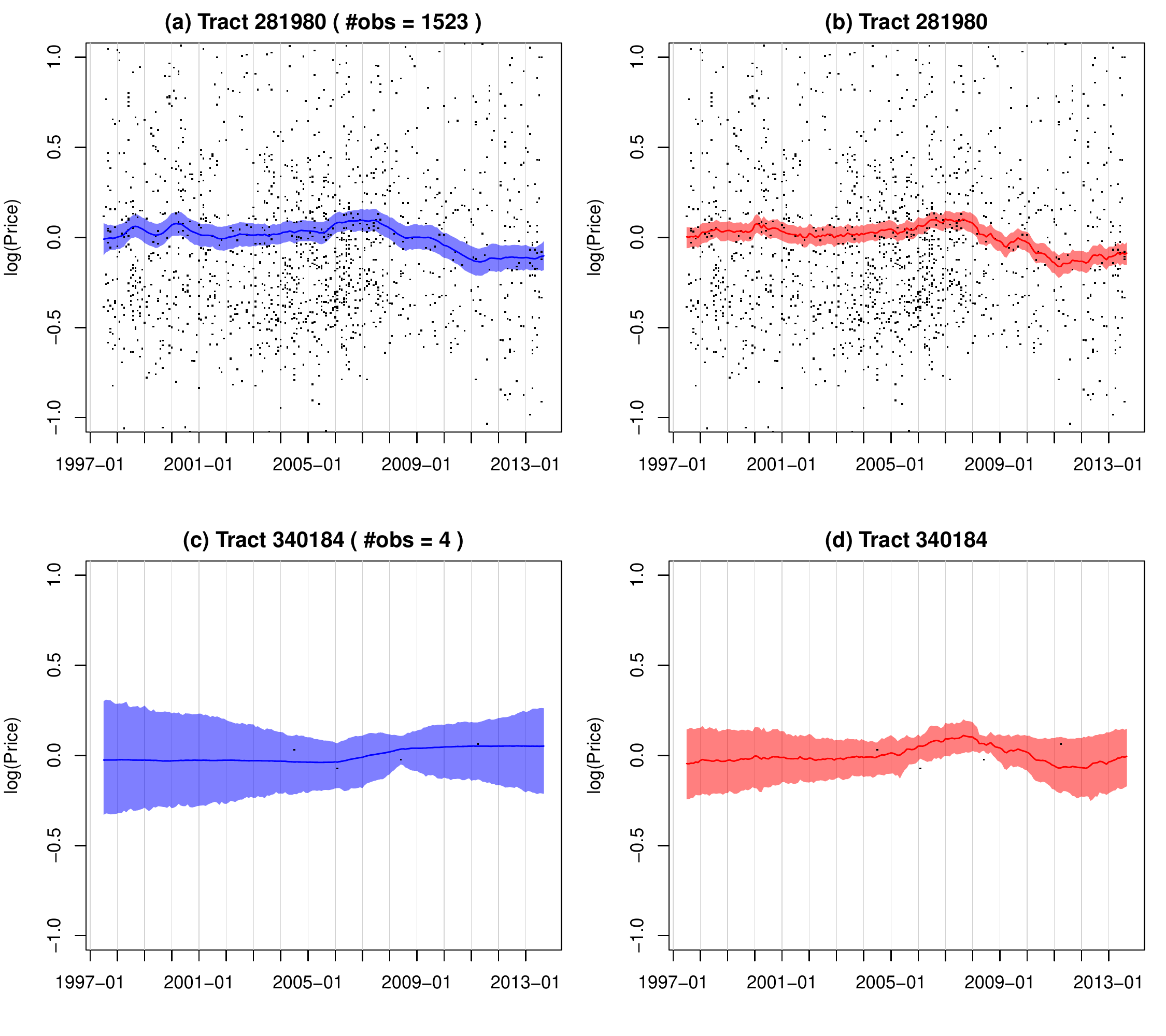}
\caption{A demonstration of the effect of clustering: (a) and (b) show the posterior mean (\emph{solid line}) and 95\% intervals (\emph{shaded gray}) for the latent price dynamics of a randomly sampled census tract with abundant observations (\emph{dots}), whereas (c) and (d) examine a tract with sparse observations.  Results are shown for models that either treat census tracts independently (\emph{left}) or allow our Bayesian nonparametric clustering of tracts with similar dynamics (\emph{right}) leading to narrower intervals, especially for tracts with few observations.}
\label{fig:effectOfClustering}
\end{figure}

The approach taken offers several advantages over existing methods.  
 Our hierarchical Bayesian nonparametric model efficiently shares information between clustered series--–a critical feature to attain high resolution.  In particular, our approach provides a form of multiple shrinkage, improving stability of our estimates in this data-scarce scenario.  
We illustrate the impact of this multiple shrinkage in Figure~\ref{fig:effectOfClustering}, with a full analysis provided in Section~\ref{sec9:realData}.
Likewise, the joint Bayesian framework considers all uncertainties together in the clustering, latent price inference and model parameter estimation.

Our paper is organized as follows. Section~\ref{sec2:data} introduces the house transaction data used in our analysis. Section~\ref{sec3:likelihood} describes the dynamical model for each census tract individually, and then the correlation structure introduced to couple the tract dynamics within a large geographic region. Section~\ref{sec4:prior} explains the prior distributions for each component in the Bayesian model. Section~\ref{sec5:modelOverview} provides a model overview and Section~\ref{sec6:posterior} proves an outline of the posterior sampling steps. Section~\ref{sec: ch7Computing} discusses some of the computational challenges and a strategy to implement the algorithm in parallel. A simulation study is provided in Section~\ref{sec8:simulation} and a detailed analysis on our Seattle housing dataset is in Section~\ref{sec9:realData}. 

%

\section{House Transaction Data}\label{sec2:data}

Our house sales data consists of 124,480 transactions in 140 census tracts of the City of Seattle from July 1997 to September 2013. Foreclosure sales are not included. For each house sale, we have the jurisdiction of the house (i.e., census tract FIPS code, zip code), month and year of the sale, the sales price, and house covariates; the latter are commonly referred to as \emph{hedonics} in the housing literature. Our hedonic variables include \texttt{number of bathrooms}, \texttt{finished square feet}, and \texttt{square feet of the lot size}. Naively, the number of bedrooms might be considered a strong predictor of the sales price.  Indeed, there is a positive correlation between the number of bedrooms and sales price; however, this can be attributed to the association of this hedonic with the total finished square feet of a home.  With the inclusion of square feet in the model, the number of bedrooms is not a significant variable.  This is common facet of many house price regression models.

\begin{figure}[t!]
\centering
\begin{tabular}{@{\hspace{0mm}}c@{\hspace{1mm}}c@{\hspace{1mm}}c@{\hspace{1mm}}c@{\hspace{1mm}}c@{\hspace{1mm}}c@{\hspace{1mm}}c}
\begin{tabular}{@{\hspace{0mm}}c@{\hspace{1mm}}c@{\hspace{1mm}}c@{\hspace{1mm}}c@{\hspace{1mm}}c@{\hspace{1mm}}c@{\hspace{1mm}}c}
\includegraphics[width=0.49\linewidth]{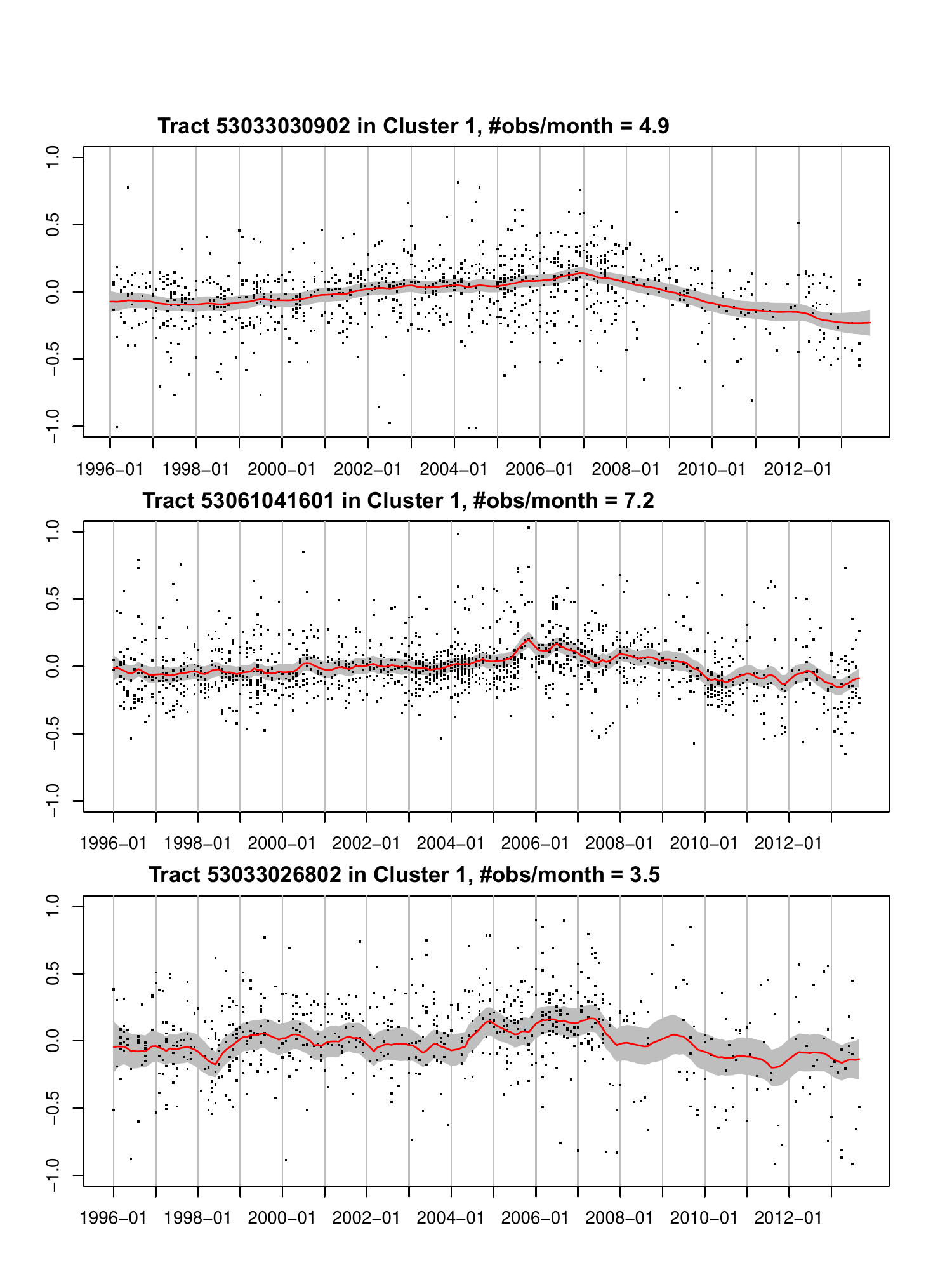}&
\includegraphics[width=0.49\linewidth]{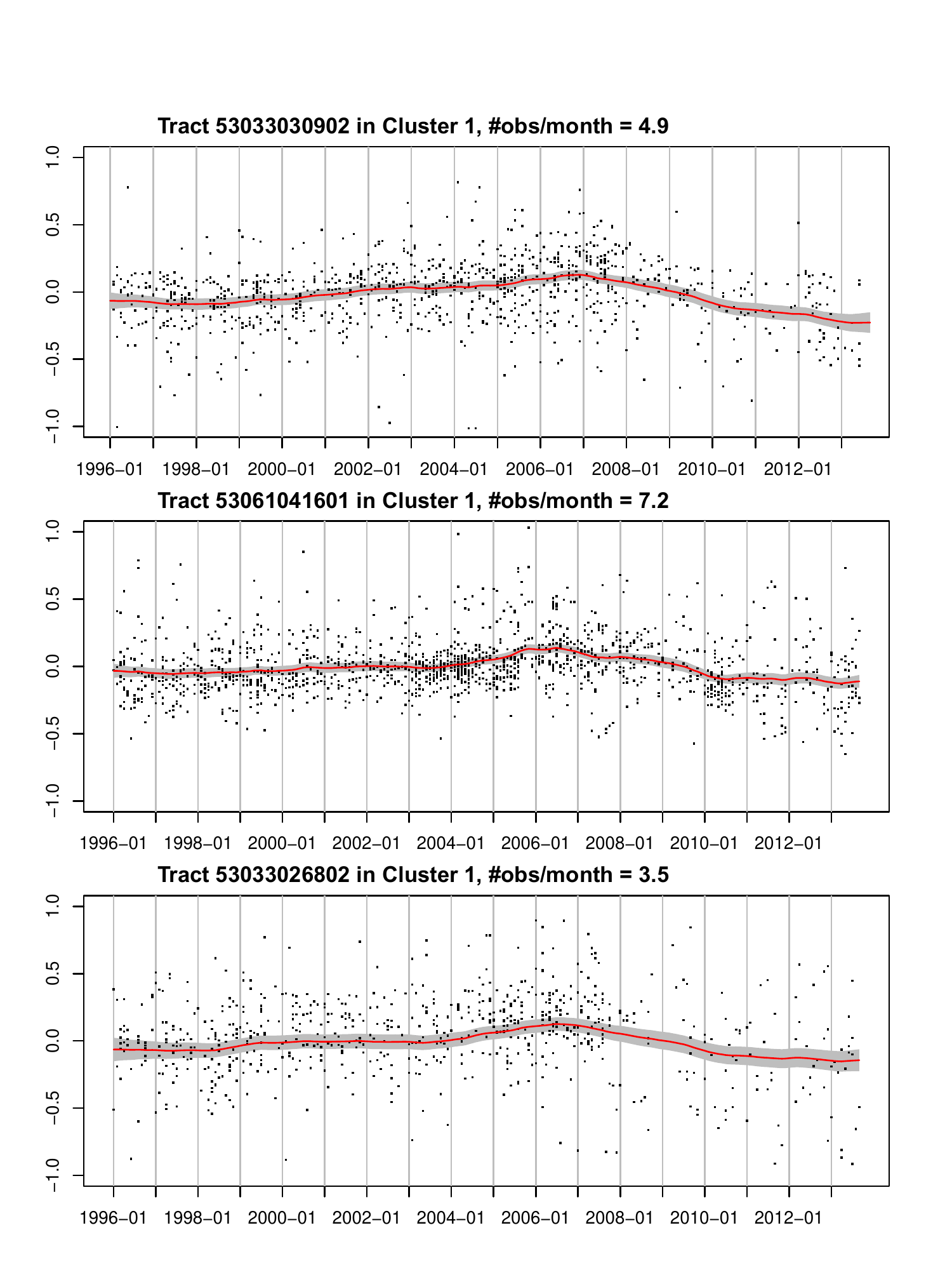}\\
(a) & (b)
\end{tabular}\\
\end{tabular}
\caption{An illustration of relating time series: (a) univariate Kalman smoother applied independently to the time series of each census tract, (b) multivariate Kalman smooth applied jointly to the tracts in the same cluster inferred using hierarchical clustering.}
\label{fig:univ_multiv_Kalman}
\end{figure}

To assess the importance of considering related regions jointly, we performed the following data analysis.  First, we examined the state space model of Eqs.~\eqref{eqn:simpleMod1}-\eqref{eqn:simpleMod2} independently across regions $i$ (whereas in Section~\ref{sec3:likelihood} the focus is on joint modeling of regions).  In that model, the latent state sequence represents the underlying price evolution of a given region---our desired index---and the observations are the individual house sales.  To infer the latent state sequence jointly with the model parameters, we use a Kalman smoother embedded in an expectation maximization (EM) algorithm.  For this analysis and that of the remainder of the paper, our spatial granularity of interest is a census tract.  We compare the performance of this independent model to one that jointly analyzes related tracts, where relatedness is determined by a hierarchical clustering approach.  The hierarchical clustering is based on $L_2$ distance between the independently Kalman smoothed estimates of the latent state sequence.  After performing the hierarchical clustering and cutting the tree at a certain level, we consider a multivariate latent state model as in Eq.~\eqref{eqn:simpleMod1} where all tracts $i$ falling in the same cluster have correlated innovations, $\epsilon_{t,i}$.  That is, $\epsilon_t^{(k)} \sim N(0,\Sigma^{(k)})$ for $\Sigma^{(k)}$ full, where $\epsilon_t^{(k)}$ is the vector of $\epsilon_{t,i}$ for tracts $i$ in cluster $k$.  The observation model remains as in Eq.~\eqref{eqn:simpleMod2}. We then applied a Kalman-smoother-within-EM algorithm to the resulting multivariate state space model.

Unsurprisingly, without sharing observations from similar tracts, the baseline independent approach does not perform well when the observations are sparse, as shown in Figure~\ref{fig:univ_multiv_Kalman}(a). In contrast, by pooling observations from other tracts, the hierarchical clustering-based latent price dynamics are smoother and with narrower intervals, as shown in Figure~\ref{fig:univ_multiv_Kalman}(b). Although this exploratory analysis motivates the importance of considering related tracts jointly, the hierarchical clustering approach considered in this section is ad-hoc since it divides the clustering and estimation into three stages rather one a unified framework.  For example, errors in the independent state estimation stage can propagate to the clusterings inferred at the second stage, which are used for the multivariate analysis in the third stage.  Additionally, the proposed multivariate model does not scale well to large clusters due to the associated large number of parameters represented by $\Sigma^{(k)}$.  In Figure~\ref{fig:univ_multiv_Kalman}(b), we simply consider a cluster with 3 tracts.  Moreover, the approach requires the user to specify the number of clusters (tree level) and distance metric used in the hierarchical clustering.  Regardless, the insights and intuition from this exploratory analysis---clustering and correlating time series---motivates the unified statistical model for relating multiple time series presented in Section \ref{sec3:likelihood}.

\section{A Model for Relating Multiple Time Series} \label{sec3:likelihood}

Recall our goal of inferring a housing valuation index for a small geographic region, e.g. census tract.  
We are faced with a large number of geographic regions and a relatively small number of observations for each region.  
Our modeling strategy is to discover price dynamics shared between these region-specific data streams, allowing us to leverage observations from related regions. 

We first describe a model for the individual housing valuation indices and then describe a clustering-based framework for correlating the processes  that share similar price dynamics. Throughout, we will assume that our geographic unit of interest is a census tract.
\subsection{Per-Series Dynamics}
We model the dynamics of the house sales prices within a census tract via a state space model.  Each census tract $i$ may have multiple house sale observations $\tilde{y}_{t,i,l}$ at time $t$. We assume that these sales are noisy, independent observations of the latent census tract value $\tilde{x}_{t,i}$ after accounting for house-level hedonics $U_\ell$ (e.g., square feet):
\begin{eqnarray}
\tilde{x}_{t,i}  &=& g_t + a_i (\tilde{x}_{t-1, i} - g_{t-1}) + \epsilon_{t,i} \quad  \epsilon_{t,i} \sim \mathcal{N}(0, \sigma_i^2)
 \label{eqn:fullMod1}\\
\tilde{y}_{t,i,l}&=& \tilde{x}_{t,i} + f_i(U_l)+v_{t,i,l} \quad v_{t,i,l} \sim \mathcal{N}(0, R_i). \label{eqn:fullMod2}
\end{eqnarray}
Our discrete-time model is indexed monthly and $g_t$ is the global market trend that captures overall, non-stationary behavior of the time series. 
To account for the hedonics, we use a census tract-specific regression $f_i(\cdot)$. 

\begin{figure}[t!]
\centering
\includegraphics[width=0.5\linewidth]{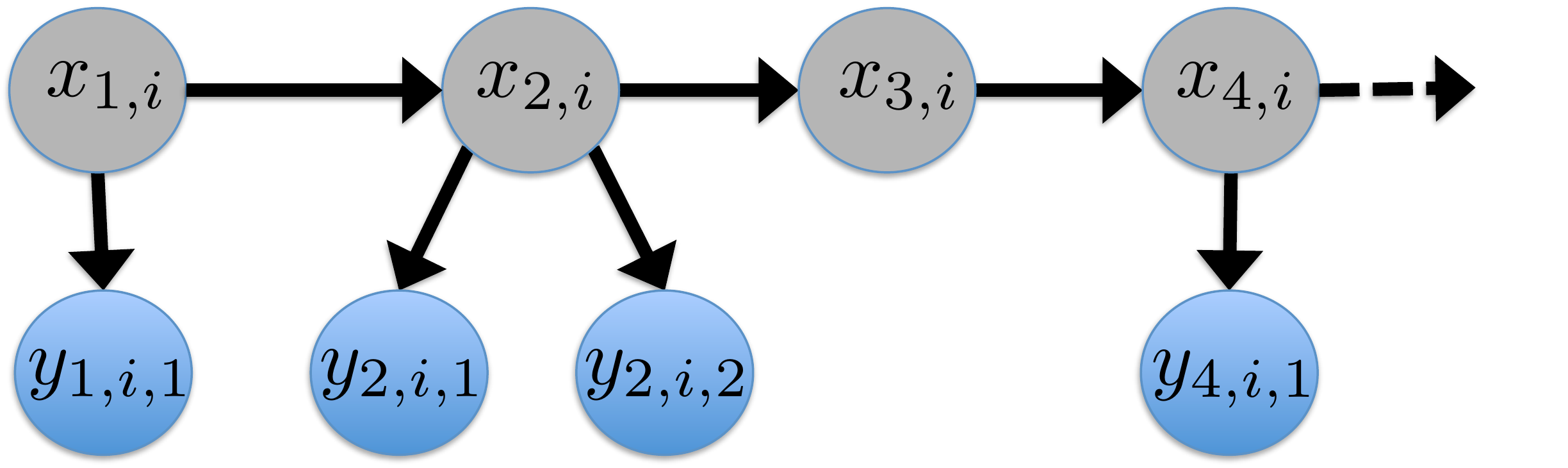}
\caption{An illustration of the state space model of Eqs.~\eqref{eqn:simpleMod1}-\eqref{eqn:simpleMod2} for census tract $i$'s data stream. The observed sales prices (after removing the global market trend) are denoted by $y_{t,i,l}$ and the (detrended) intrinsic price of census tract $i$ by $x_{t,i}$. }
\label{fig:HMM}
\end{figure}

For the sake of simplicity, since our focus is on small geographic regions,
we assume that the global trend $g_t$ is known or pre-calculated based on all transactions in the market. 
Computing a global trend is relatively straightforward since we have sufficient data in aggregate.
Instead, we focus on modeling the deviance of the latent dynamics of census tracts from the global market trend. 
This deviance can be defined as $x_{t,i} \equiv \tilde{x}_{t,i} - g_t$. To further simplify the model, we assume the house feature function $f_i(\cdot)$ is composed of linear basis functions. The simplified model is
\begin{eqnarray}
x_{t,i}  &=& a_i x_{t-1, i} + \epsilon_{t,i} \quad  \epsilon_{t,i} \sim \mathcal{N}(0, \sigma_i^2)\label{eqn:simpleMod1}\\
y_{t,i,l} &=&  x_{t,i} + \sum_{h=1}^H \beta_{i,h} U_{l,h}+v_{t,i,l} \quad v_{t,i,l} \sim \mathcal{N}(0, R_i) \; , \label{eqn:simpleMod2}
\end{eqnarray}
where $y_{t,i,l} \equiv \tilde{y}_{t,i,l} - g_t$ represents the deviance of each house sales price at time $t$ from the global market price at that time. We refer to the latent $x_{t,i}$ order 1 autoregressive process (AR(1)) in Eq.~\eqref{eqn:simpleMod1} as the \emph{intrinsic price dynamics} for each census tract. Since we are modeling deviances from the global trend, the choice of a stationary process is reasonable. We call the series of observations from one census tract a \emph{data stream}. Eqs. (\ref{eqn:simpleMod1}) and (\ref{eqn:simpleMod2}) are akin to a standard linear-Gaussian state space model, but with a varying number (potentially 0) of observations $y_{t,i,l}$ of a given state $x_{t,i}$, as illustrated in Figure~\ref{fig:HMM}.

%

\subsection{Relating Multiple Data Streams}\label{sec:correlatingTS}
There are clearly temporal trends to house values, and these trends may vary significantly and sometimes abruptly across geographic locations. 
We want to share information between related tracts which exhibit similar temporal trends, and aim to discover these groups of tracts from the observed data streams. 
The idea for clustering multiple data streams has two justifications.  From a data generating perspective, housing price dynamics are naturally clustered due to a number of factors, including the composition of homes, number of foreclosures, school district boundaries, crime rate, and the proximity to parks, waterfront and other amenities.
From a statistical inference perspective, clustering census tracts increases power and precision in parameter estimation by pooling the observations from grouped data streams.

We now seek to define the mechanism by which data streams relate, and then use this to cluster the series.  If house prices in one neighborhood increase, prices in related neighborhoods are also likely to increase. This type of sharing of price dynamics can be modeled by correlating the \emph{innovation} terms $\epsilon_{t,i}$ in the intrinsic price dynamics of Eq. (\ref{eqn:simpleMod1}). We then cluster all series with correlated innovations and assume independence between the dynamics of those falling into separate clusters.  More specifically, let $z_i=k$ denote that census tract $i$ is in cluster $k$ and $\boldsymbol{\epsilon}_t^{(k)}$ be the vector of innovations $\epsilon_{t,i}$ for census tracts in cluster $k$.  Instead of treating the $\epsilon_{t,i}$ independently across $i$, the intrinsic price dynamics $x_{t,i}$ within cluster $k$ can be correlated by considering $\boldsymbol{\epsilon}_t^{(k)} \sim \mathcal{N}(0, \Sigma_k)$ for $\Sigma_k$ non-diagonal. We assume $\boldsymbol{\epsilon}_t^{(k)}$ is independent of $\boldsymbol{\epsilon}_t^{(j)}$ for all $j \neq k$.  Stacking up all $\boldsymbol{\epsilon}_t^{(k)}$, $k=1, \cdots, K$, into a large ${\boldsymbol \epsilon_t}$ vector of length $p$ (the number of census tracts), our model is equivalent to ${\boldsymbol \epsilon_t} \sim N(0,\Sigma)$ for $\Sigma$ block diagonal with blocks $\Sigma_k$.

Jointly clustering census tracts \emph{and} correlating the dynamics within a given cluster is a challenging task; it is equivalent to inferring the block structure of $\Sigma$, which entails discovering both an ordering on the census tracts and the dimensions of the blocks $\Sigma_k$.  
Since we do not assume that the number of clusters (number of blocks in $\Sigma$) is know, this adds an additional challenge.

To generatively define block diagonal covariance matrices with unknown block sizes, we leverage {\bf latent factor models}. 
We start by assuming that there are $K$ clusters with known membership, and then revisit the idea of inferring the memberships and number of clusters in Section~\ref{sec4:prior}. In particular, consider:
\begin{eqnarray}
\epsilon_{t,i} = \lambda_{iz_i} \eta^*_{t,z_i} + \tilde{\epsilon}_{t,i} \quad \tilde{\epsilon}_{t,i} \sim N(0, \sigma_0^2) \quad \eta^*_{t,k} \sim N(0,1).  \label{eqn:simpleMod3}
\end{eqnarray}
Here, $\eta^*_{t,k}$ is the latent factor associated with cluster $k$ at time $t$, $\lambda_{ik}$ is the factor loading for census tract $i$ assuming it is in cluster $k$, and $\tilde{\epsilon}_{t,i}$ is idiosyncratic noise drawn independently over time and tracts. 
We model $\lambda_{ik} \sim N({\mu_{\lambda}} , \sigma_{\lambda}^2)$.  
We can then write ${\boldsymbol \epsilon_t} = (\Lambda \cdot Z) {\boldsymbol \eta^*_t} + \tilde{\boldsymbol \epsilon}_t$, where $\Lambda$ is a $p \times K$ Gaussian matrix, $Z$ is an indicator matrix with $Z_{ik} = {1}[z_i = k], \; {\boldsymbol \eta}^*_t \sim \mathcal{N}_K(0, I)$, and $\tilde{\boldsymbol \epsilon}_t \sim \mathcal{N}_p(0, \sigma^2_0 I)$. Here, $A \cdot B$ represents the element-wise product. Conditioned on the factor loading matrices $\Lambda$ and $Z$, the covariance for ${\boldsymbol \epsilon}_t$ is $\Sigma = (\Lambda \cdot Z)(\Lambda \cdot Z)^T + \sigma_0^2 I$. Equivalently,
\begin{eqnarray}
\textrm{cov}(\epsilon_{t,i}, \epsilon_{t,i'}| \Lambda, Z) = \left\{ \begin{array}{ll} \lambda_{ik} \lambda_{i'k} + \sigma_0^2 \delta(i,i') & z_i = z_{i'}=k, \forall k \\ 0 & \textrm{otherwise}. \end{array} \right. \label{eqn:Qstructure}
\end{eqnarray} 
From Eq. (\ref{eqn:Qstructure}), the conditional covariance for ${\boldsymbol \epsilon_t}$ is a block-diagonal matrix defined by the clusterings specified by $z_i$. That is, data streams within the same cluster will have correlated dynamics, and those in different clusters will evolve independently. 

To infer the clustering of region-specific data streams, we propose a Bayesian nonparametric approach using a Dirichlet Process (DP) prior that leads to an adaptive, data-driven clustering, allowing for an unknown number of blocks (clusters) in the covariance. This model is related to that of  \cite{nonparaClusterNIPS2012}, but specified for the time series domain.  The details of our prior specification are in Section~\ref{sec4:prior}.

\section{Prior Specification}\label{sec4:prior}
In this section, we describe the prior specifications for our various model parameters.

\subsection{Cluster Membership} We first provide background on the DP in Section~\ref{sec:DP}, and then in Section~\ref{sec:DPLFM} describe how we utilize this prior in our dynamical model to cluster tracts with correlated intrinsic price dynamics in the presence of an unknown number of clusters. 

\subsubsection{The Dirichlet Process} {\label{sec:DP}}
A DP \citep{Blackwell1973,Ferguson1973_BayesNonparaDPMM} is a distribution over countably infinite discrete probability measures. A draw $G \sim DP(\alpha, G_0)$, with concentration parameter $\alpha$ and base measure $G_0$, can be constructed as 
\begin{eqnarray}
G=\sum_{k=1}^{\infty} \pi_k \delta_{\theta_k^*}, \quad \theta_k^* \sim G_0,
\end{eqnarray}
where the mixture weights $\pi_k$ are sampled via a stick breaking construction \citep{stickBreaking1994}:
\begin{eqnarray}
\pi_k = v_k \prod_{j=1}^{k-1} (1-v_j), \quad v_k \sim \textrm{Beta}(1, \alpha).
\end{eqnarray}
We denote the stick breaking process as ${\boldsymbol{\pi}} \sim $ GEM$(\alpha)$, where $\boldsymbol{\pi}=(\pi_1, \pi_2, \cdots)$.
The DP prior produces clusters of $\theta_i\sim G$, $i=1,\cdots, n$, due to the fact that $G$ is a discrete probability measure (i.e., multiple $\theta_i$ are sampled with identical values $\theta^*_k$). 
Equivalently, we can introduce cluster indicators $z_i \sim \pi$ such that $z_i=k$ implies that $\theta_i$ takes the unique value $\theta_k^*$.  That is, $\theta_i = \theta_{z_i}^*$. 

Integrating out the stick breaking measure ${\boldsymbol{\pi}}$, the predictive distribution of $z_i$ given the memberships of other tracts ${\bf z}_{-i}$ is  
\begin{eqnarray}
P(z_i=k|{\bf z}_{-i}, \alpha) \propto  \left\{ \begin{array}{ll}
 \frac{n_{-i,k}}{n-1+\alpha} & \textrm{ for } k=1,\cdots,K\\
 \frac{\alpha}{n-1+\alpha} & \textrm{ for } k=K+1,
\end{array} \right. 
\label{eqn:CRPcollapsed}
\end{eqnarray}
where $K$ indicates the number of unique values of $z_i$ in ${\bf z}_{-i}$. That is, tract $i$ may join one of the existing clusters with probability proportional to the size of the cluster, $n_{-i,k}$, or start a new cluster with probability proportional to $\alpha$. The resulting sequence of partitions is referred to as the {\em Chinese Restaurant Process} (CRP)~\citep{CRP_Pitman2006}. 

\subsubsection{Clustering of Regions by Latent Dynamic Factors}\label{sec:DPLFM}
In our housing application, we place a DP prior on the parameter by which we wish to cluster the census tract intrinsic dynamics.  As detailed in Section~\ref{sec3:likelihood}, we relate the data streams within a cluster via correlated dynamics induced by a latent factor model with a cluster-specific latent factor process $\eta_{1:T,k}^*$.  As such, to specify a Bayesian nonparametric clustering model we take $\eta_{1:T,i} \sim G$ with $G \sim \mbox{DP}(\alpha,G_0)$, where $\eta_{1:T,i}$ is the latent factor process for census tract $i$.  In our indicator variable representation, we define mixture weights $\pi \sim \mbox{GEM}(\alpha)$, cluster-specific parameters $\eta_{1:T,k}^* \sim G_0$, and cluster indicators $z_i \sim \pi$ such that $\eta_{1:T,i} = \eta_{1:T,z_i}^*$.  That is, $\eta_{1:T,i}$ serves the role of $\theta_i$ and $\eta_{1:T,k}^*$ equates with $\theta_k^*$ in the generic Dirichlet process mixture model of Section \ref{sec:DP}.  The base measure $G_0$ is specified as a multivariate normal distribution $\mathcal{N}_T(0, I)$ such that $\eta^*_{t,k} \sim N(0,1)$ for $t=1,\dots, T$, $k=1,2,\dots$. 

\subsection{Latent Autoregressive Process Parameters}
The latent autoregressive (AR) process in Eq. (\ref{eqn:simpleMod1}) has an autoregressive parameter $a_i$, a factor loading $\lambda_{ik}$, and the variance of the idiosyncratic noise $\sigma_0^2$. We place conjugate priors on these parameters, respectively:
\begin{eqnarray}
a_i &\sim& \mathcal{N} (\mu_a, \sigma_a^2), \quad i=1,\dots, p\\
\lambda_{ik} &\sim& \mathcal{N}(\mu_{\lambda}, \sigma_{\lambda}^2), \quad i=1, \cdots, p, \,\, k=1,2,\dots\\
\sigma_0^2 &\sim& IG(\alpha_{\epsilon 0}, \beta_{\epsilon 0}).
\end{eqnarray}
The hyperparameters $\mu_a,\sigma_a^2,\mu_{\lambda}, \sigma_{\lambda}^2$ are given priors as well.
Details of these hyperpriors and settings of the hyperparameters $\alpha_{\epsilon 0}, \beta_{\epsilon 0}$ are provided in Supplement \ref{App:hyperPriorSigma0Squared}.

\subsection{Emission Parameters} Recalling the emission process in Eq. (\ref{eqn:simpleMod2}), we place conjugate priors on the tract-specific hedonic parameters $\beta_{i,h}$ and observational variance $R_i$:  
\begin{eqnarray}
\beta_{i,h} &\sim& \mathcal{N}(\mu_h, \sigma_h^2) \quad i=1,\cdots, p, \,\, h=1,\cdots, H \\
R_i &\sim& IG(\alpha_{R0}, \beta_{R0}) \quad i=1,\cdots, p. 
\end{eqnarray}
We further assume priors on $\mu_h$ and $\sigma_h^2$.
These hyperpriors and the values of the hyperparameters $\alpha_{R0}$ and $\beta_{R0}$ are provided in Supplement \ref{App:hyperPriorAlphaBetaR0}. 

\section{Model Overview}\label{sec5:modelOverview}

\begin{figure}[t!]
\centering
\includegraphics[width=1\linewidth]{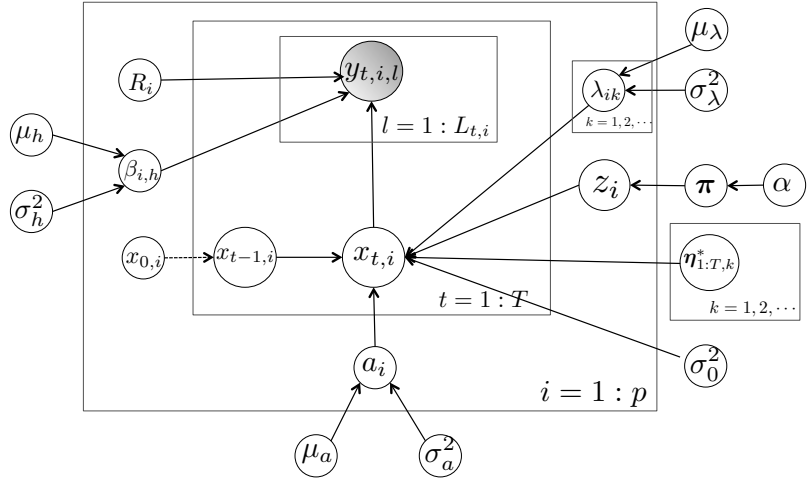}
\vspace{-0.2in}
\caption{Graphical model representation of our Bayesian nonparametric house sales dynamic model summarized in Section~\ref{sec5:modelOverview}. Boxes indicate replication of random variables and shaded nodes the observations.  Note that $x_{1:T}$ forms a length $T$ Markov chain realization; our box here is an abuse of notation used for compactness.}
\label{fig:graphicalModel}
\end{figure}

Our model assumes that the observed house sales prices center about the intrinsic price of the associated census tract and transaction month, after accounting for hedonic effects. The intrinsic price for each census tract follows an AR(1) marginally. The DP provides a flexible prior for nonparametric clustering of the intrinsic price dynamics associated with each census tract based on our latent factor model, which induces correlation of price dynamics within a cluster. Figure~\ref{fig:graphicalModel} shows the graphical model representation. The overall model specification for the Bayesian nonparametric house sale dynamic model is summarized as:
\begin{enumerate}
\item Draw Dirichlet process realization $G\sim DP(\alpha, G_0):$
\begin{eqnarray}
G = \sum_{k=1}^{\infty} \pi_k \delta_{\theta_k^*}, \quad \textrm{ where } \theta_k^*= {\boldsymbol \eta}^*_{1:T,k}
\end{eqnarray}
\item For the data stream associated with each census tract $i$ from $1$ to $p$:
\begin{enumerate}
\item Draw cluster membership $z_i|{\boldsymbol \pi} \sim {\boldsymbol \pi}$
\item Draw factor loadings $\lambda_{ik} \sim \mathcal{N}(\mu_\lambda, \sigma_\lambda^2)$
\item For each timestep $t$ from $1$ to $T$:
\begin{enumerate}
\item Draw the state sequence $x_{t,i}|x_{t-1,i},z_i \sim \mathcal{N}(a_i x_{t-1,i} + \lambda_{iz_i}\eta^*_{t,z_i}, \; \sigma_0^2)$
\item Draw an observation $y_{t,i,l}|x_{t,i} \sim \mathcal{N}\left(x_{t,i}+\sum_{h=1}^H \beta_{i,h}U_{l,h}, \; R_i\right)$
\end{enumerate}
\end{enumerate}
\end{enumerate}

\section{MCMC Posterior Computations}\label{sec6:posterior}
Our posterior computations are based on a Gibbs sampler, with steps detailed below. Scaling this sampling strategy to our large housing dataset is discussed in Section \ref{sec: ch7Computing}.

Letting $\boldsymbol{\psi}=\left\{\mathbf{a}=\{a_i\}, \boldsymbol{\lambda}=\{\lambda_{ik}\}, \mathbf{R}=\{R_i\}, \boldsymbol{\beta}=\{\beta_{i,h}\} ,\sigma_0^2\right\}$ and $\boldsymbol{\psi}^{(k)}$ the associated subset of parameters corresponding to the $k$-th cluster based on assignments $\mathbf{z}=\{z_i\}$, the Gibbs sampler is outlined as follows: 
\begin{framed}
\begin{enumerate}
\item Sample $z_i=k | \mathbf{z}_{-i}, \alpha, \mathbf{y}, \boldsymbol{\psi}$. Note we marginalize the stick-breaking random measure $\boldsymbol{\pi}$, the latent housing valuation processes $\mathbf{x}^{(k)}$, and the cluster latent factor processes $\boldsymbol{\eta}^{*(k)}$.\label{sec6.1:sampleZ}
\item Impute $\mathbf{x}$ and $\boldsymbol{\eta}^*$ as auxiliary variables.  Specifically, block sample $\mathbf{x}$, $\boldsymbol{\eta}^*$ as $\mathbf{x}^{(k)} | \mathbf{z}, \mathbf{y}^{(k)}, \boldsymbol{\psi}^{(k)}$ and $\boldsymbol{\eta}^* | \mathbf{z}, \mathbf{x}, \boldsymbol{\psi} $.

\item Sample $\boldsymbol{\psi}^{(k)}|\mathbf{z},\mathbf{y}^{(k)}, \mathbf{x}^{(k)}, \boldsymbol{\eta}^{*(k)}$

\item Discard  $\mathbf{x}$ and $\boldsymbol{\eta}^*$ to sample hyperparameters conditional on $\boldsymbol{\psi},\mathbf{z}$.
\end{enumerate}
\end{framed}

\subsection{Sampling the cluster membership}\label{sec:CRPsampling}
We sample the cluster indicators $z_{i}$ conditional on model parameters and house sales transactions. We analytically marginalize out the infinite set of mixture weights ${\boldsymbol \pi}$, latent factor process $\boldsymbol{\eta}^*=\{\eta_{1:T,k}^*\}$ and the intrinsic dynamics $\mathbf{x}=\{x_{1:T,i}\}$. Specifically, the full conditional of indicator $z_i$ for census tract $i$ is:
\begin{multline}
P\left(z_i=k \left|{\bf z}_{-i}, {\bf y}_{1:T}, \{a_j\}^{(k)},a_i,\{\lambda_{jk}\},\sigma_0^2,\{R_j\}^{(k)},R_i, \{\beta_{j,h}\}^{(k)},\{\beta_{i,h}\}, \alpha \right.\right) \\ \propto 
 P\left(z_i=k \left|{\bf z}_{-i}, \alpha \right.\right)
P\left({\bf y}_{1:T,i} \left| z_i=k,\mathbf{z}_{-i}, {\bf y}_{1:T,-i}^{(k)}, \{a_j\}^{(k)},\Sigma^{(k)},\{R_j\}^{(k)},\{\beta_{j,h}\}^{(k)} \right.\right). 
\label{eqn:sampleZ}
\end{multline}
The first factor is the prior belief of cluster membership for tract $i$ conditional on memberships of all other tracts, which results from the CRP prior of Eq. ({\ref{eqn:CRPcollapsed}}) (and the use of exchangeability). The second factor is the likelihood of the data stream for tract $i$ assuming membership to cluster $k$.  The marginalization over $\mathbf{x}$ and $\boldsymbol{\eta}^*$ results in a dependence upon all other data streams in cluster $k$, ${\bf y}_{1:T,-i}^{(k)}$, and the covariance between intrinsic dynamics in the cluster, $\Sigma^{(k)}$, specified via Eq. (\ref{eqn:Qstructure}).  The other model parameters for cluster $k$ include: the AR coefficients $\{a_j\}$, observational variances $\{R_j\}$, and hedonic effects $\{\beta_{j,h}\}$ for all tracts $j$ in cluster $k$ (including $i$ when conditioning upon $z_i=k$).  We denote these restricted sets via $\{\cdot\}^{(k)}$.

A message passing scheme along the entire sequence of length $T$ is required to compute the likelihood of the $i$th data stream conditioned on all others in cluster $k$, integrating over the intrinsic dynamics $\mathbf{x}_{1:T}^{(k)}$. This algorithm is essentially a Kalman filter, but allows for a varying number of observations per time step, including no observations for some time periods. The detailed algorithm is provided in Supplement \ref{AppSub:condLike_allObs}. 

For the special case of census tract $i$ creating a new cluster, i.e. $z_i = K+1$, the prior belief follows the CRP prior of Eq. ({\ref{eqn:CRPcollapsed}}). The likelihood becomes simply $P\left({\bf y}_{1:T,i} \left| \mathbf{z}, a_i,\Sigma^{(K+1)},R_i,\beta_{i,h} \right.\right)$, where $\Sigma^{(K+1)} = \lambda_{i,K+1}^2+\sigma_0^2$, as specified in Eq. (\ref{eqn:Qstructure}) and having sampled $\lambda_{i,K+1} \sim \mathcal{N}(\mu_{\lambda}, \sigma_{\lambda}^2)$ for all tracts, but marginalizing $\eta^*_{1:T,K+1}$.  This represents a variant of Neal's Algorithm 8 for sampling from DP models~\citep{Neal2000}.

\subsection{Block-sampling the intrinsic price dynamics ${\mathbf x}$ and cluster latent factor processes ${\boldsymbol \eta}^*$} 
To block sample $(\mathbf{x},\boldsymbol{\eta}^*)$, we first sample the intrinsic price dynamics ${\mathbf x}_{1:T}^{(k)}$ jointly for all tracts in cluster $k$, analytically marginalizing $\boldsymbol{\eta}^*$.  To do this, we use a forward filter backward sampler (FFBS) outlined in Supplement \ref{App:sampleX}.

We then sample $\boldsymbol{\eta}^*$ given $\mathbf{x}$.  Recall that the intrinsic price dynamics for multiple tracts in the same cluster $k$ are correlated through the common latent factor process $\eta^*_{1:T,k}$ for the AR(1) innovations $\boldsymbol{\epsilon}^{(k)}$, as in Eq. (\ref{eqn:simpleMod3}). By conjugacy, we sample the cluster-specific latent factor $\eta^*_{t,k}$ for time period $t=1, \cdots, T$ and $K$ existing clusters as follows:
{\small
\begin{eqnarray}
\qquad {\boldsymbol \eta}^*_t \left| {\boldsymbol \lambda},{\bf z,x}, {\bf a}, \sigma_0^2 \right. 
\sim  \mathcal{N}_K
\left\{
\begin{array}{cc}
V\frac{1}{\sigma_0^2}(\Lambda \cdot Z)^T ({\bf x_t}-A{\bf x_{t-1}}) ,\\
V = \left[ I_K + \frac{1}{\sigma_0^2}(\Lambda \cdot Z)^T(\Lambda \cdot Z) \right]^{-1} 
\end{array} 
\right\}. 
\end{eqnarray}}
The derivation is provided in Supplement \ref{App:sampleEta}.

\subsection{Sampling factor loadings $\lambda$}
We sample the loadings $\lambda_{ik}$ of the loadings matrix $\Lambda_{p \times K}$ as
\begin{eqnarray}
\lambda_{ik}|{\bf x, a, \boldsymbol{\eta}^*, z}, \sigma_0^2 \sim \mathcal{N}(\mu_{ik}^*, v_{ik}^*),
\end{eqnarray}
where 
\begin{eqnarray}
(\mu_{ik}^*, v_{ik}^*)=\left\{ \begin{array}{ll}
\mu_\lambda, \sigma_\lambda^2 & \textrm{if } Z_{ik}=0\\
v\left(\frac{\mu_\lambda}{\sigma_\lambda^2}+\frac{1}{\sigma_0^2} \sum_{t=1}^T \epsilon_{t,i}\eta^*_{t,k}\right),
v=\left( \frac{1}{\sigma_\lambda^2}+\frac{1}{\sigma_0^2} \sum_{t=1}^T (\eta^*_{t,k})^2 \right)^{-1} & \textrm{if } Z_{ik}=1
\end{array} \right. .  \notag
 \label{eqn:lambdaStar}
\end{eqnarray}
Here, we recall the definition of the membership matrix $Z$ from Section~\ref{sec:correlatingTS}. Note that $\epsilon_{t,i}=x_{t,i}-a_ix_{t-1,i}$ and $\sum_{t=1}^T \epsilon_{t,i}\eta^*_{t,k}$ can be written as the inner product ${\boldsymbol{\epsilon}_i^T \boldsymbol{\eta}^{*(k)}}$. The derivation is given in Supplement \ref{App:sampleLambda}.

\subsection{Sampling AR parameters $a_i$} 


Using conjugacy results of the normal distribution, and conditioning upon a cluster assignment $z_i=k$, we sample the tract-specific AR coefficient $a_i$ for $i= 1, \cdots, p$ as
\begin{multline}
a_i \left|z_i=k,{\mathbf x}_i, \boldsymbol{\eta}^{*(k)}, \lambda_{ik},\sigma_0^2, \mu_a, \sigma_a^2 \right. \\
\sim \mathcal{N}\left( V\left[\frac{\mu_a}{\sigma_a^2}  
+ \sum_{t=1}^T \left(\frac{x_{t-1,i}^2}{\sigma_0^2} \cdot \frac{x_{t,i}-\lambda_{ik}\eta^*_{t,k}}{x_{t-1,i}} \right) \right], 
V=\left( \frac{1}{\sigma_a^2}+\sum_{t=1}^T \frac{x_{t-1,i}^2}{\sigma_0^2} \right)^{-1} \right) \notag.
\end{multline}
The derivation is provided in Supplement \ref{App:SampleE}.

\subsection{Sampling emission parameters $R, \beta, \sigma_0^2$} 
By conjugacy, we can sample the observation variance $R_i$ for $i= 1, \cdots, p$ as
\begin{align}
R_i \left|{\bf x}_{1:T,i}, {\bf y}_{1:T,i}, \alpha_{R0}, \beta_{R0} \right.
\sim \mbox{IG} \left(\alpha_{R0}+\frac{1}{2} m_i, \, \beta_{R0}+\frac{1}{2}\sum_{t=1}^{T}\sum_{l=1}^{Lt}(y_{t,i,l}-x_{t,i})^2 \right), \notag
\end{align}
where $m_i$ is the number of transactions in census tract $i$. The values
of the hyperparameters $\alpha_{R0},\beta_{R0}$ are provided in Supplement \ref{App:hyperPriorAlphaBetaR0}.

We sample the covariate effect $\beta_{i,h}$ for $i=1, \cdots,p$ and $h=1,\cdots, H$ as
\begin{multline}
\beta_{i,h} | \mu_h, \sigma_h^2, R_i, {\bf x}_{1:T,i}, {\bf y}_{1:T,i} \\
\sim N
\left\{ 
\begin{array}{cc}
v\left[  \frac{\mu_h}{\sigma_h^2} + \frac{1}{R_i} \sum_{t=1}^T\sum_{l=1}^{L_t}U_{l,h} \left(y_{t,i,l}-x_{t,i}-\sum_{s\neq h}\beta_{i,s}U_{l,s}\right) \right] , \\
v = \left(\frac{1}{\sigma_h^2} + \frac{1}{R_i}\sum_{t=1}^T\sum_{l=1}^{L_t}U_{l,h}^2\right)^{-1}
\end{array}
\right\}. \notag
\end{multline}

Finally, the variance parameter $\sigma_0^2$ has full conditional
\begin{multline}
\sigma_0^2|{\boldsymbol \lambda}, {\boldsymbol \eta}^*, {\bf a,z,x}, \alpha_{\epsilon 0}, \beta_{\epsilon 0} \\
\sim \mbox{IG}\left( \alpha_{\epsilon 0} + \frac{Tp}{2} , \beta_{\epsilon 0}+\frac{1}{2}
\sum_{t=1}^T\sum_{i=1}^n\sum_{k=1}^K z_{ik}(x_{t,i}-a_ix_{t-1,i}-\lambda_{ik}\eta^*_{t,k})^2 \right). \notag
\end{multline}
The details can be found in Supplement \ref{App:SampleF}.

\subsection{Sampling hyperparameters}
The hyperparameters $\mu_\lambda,\sigma^2_\lambda,\mu_a,\sigma^2_a$ and $\mu_h,\sigma_h^2$ for hedonics $h=1,\dots,H$ can be sampled as follows:

\begin{align}
	\mu_\lambda \left|{\bf z}, \boldsymbol{\lambda}, \sigma_\lambda^2, \mu_{\lambda 0}, \sigma_{\lambda 0}^2 \right.
&\sim N
\left[ 
\begin{array}{cc}
v\left(\frac{\mu_{\lambda 0}}{\sigma_{\lambda 0}^2} + \frac{1}{\sigma_\lambda^2} \sum_{k=1}^K\sum_{i:z_i=k}  \lambda_{ik} \right) ,
v = \left( \frac{1}{\sigma_{\lambda 0}^2}+\frac{p}{\sigma_\lambda^2} \right)^{-1} 
\end{array}
\right] \\
\sigma_\lambda^2 \left|{\bf z}, \boldsymbol{\lambda}, \mu_\lambda, \alpha_{\lambda 0}, \beta_{\lambda 0} \right.
&\sim \mbox{IG}\left( \alpha_{\lambda 0} + \frac{p}{2}, \beta_{\lambda 0} + \frac{1}{2} \sum_{k=1}^K\sum_{i:z_i=k}  (\lambda_{ik}-\mu_\lambda)^2 \right)\nonumber
\end{align}
\begin{align}
\mu_a \left|\{a_i\}, \sigma_a^2, \mu_{a0}, \sigma_{a0}^2 \right.
&\sim
N\left[ v\left(\frac{\mu_{a0}}{\sigma_{a0}^2} + \frac{1}{\sigma_a^2}\sum_{i=1}^p a_i \right), v=\left( \frac{1}{\sigma_{a0}^2}+\frac{p}{\sigma_a^2}\right)^{-1}\right] \\
\sigma_a^2 \left| \{a_i\},\mu_a, \alpha_{a0},\beta_{a0}  \right.
&\sim IG\left[ \alpha_{a0}+\frac{p}{2}, \beta_{a0} + \frac{1}{2} \sum_{i=1}^p(a_i-\mu_a)^2\right]\nonumber
\end{align}
\begin{align}
 \mu_{h}|\beta_{1:p, h}, \sigma_h^2, \mu_{h0}, \sigma_{h0}^2 
&\sim N \left[ v\left(  \frac{\mu_{h0}}{\sigma_{h0}^2}+ \frac{1}{\sigma_h^2}\sum_{i=1}^p \beta_{i,h} \right),
v=\left( \frac{1}{\sigma_{r0}^2} + \frac{p}{\sigma_r^2} \right)^{-1} \right] \\
\sigma_h^2|\beta_{1:p,h}, \mu_h, \alpha_{h0}, \beta_{h0} 
&\sim \mbox{IG} \left[\alpha_{h0}+\frac{p}{2}, \beta_{h0}+\frac{1}{2}\sum_{i=1}^p(\beta_{i,h}-\mu_h)^2 \right].\nonumber
\end{align}

\subsection{Sampling the DP hyperparameter} 
We assume a hyperprior for the DP concentration parameter $\alpha \sim \textrm{Gamma}(\alpha_\alpha, \beta_\alpha)$ and follow the sampling procedure suggested by \cite{Escobar94bayesiandensity}. Details are provided in Supplement \ref{App:SampleG}.

\section{Computational challenges and strategies} \label{sec: ch7Computing}
Although marginalizing $\boldsymbol{\pi}, \mathbf{x},$ and $\boldsymbol{\eta}$---i.e., considering a \emph{collapsed} sampler--- reduces the dimension of the posterior we explore in our sampling, the marginalization of $\boldsymbol{\pi}$ induces dependencies between the $z_i$.  As such, we must rely on the CRP-based sequential sampling described in Section~\ref{sec:CRPsampling}.  Involved in this sampling is a computationally intensive likelihood evaluation.  In particular, for each census tract $i$ we must consider adding the tract to each existing cluster $k$, each of which involves a Kalman-filter-like algorithm.  Naively, just harnessing the Woodbury matrix identity yields a computational complexity of $O((\min\{ n^{(k)},p^{(k)}\})^3T)$, where $n^{(k)}$ is the maximum number of observations at any time $t$ aggregated over census tracts in cluster $k$ and $p^{(k)}$ is the number of census tracts in cluster $k$.  In most cases, we have $n^{(k)} >> p^{(k)}$.

To address the computational challenge of coupled $z_i$---which at first glance seems to imply reliance on single machine serial processing--- we adopt the clever trick of \cite{WilliamsonParallelMCMC2013} for parallel collapsed MCMC sampling in DP mixture models (DPMM). A similar approach was proposed by \cite{RyanAdams}. The conventional DPMM assumes that observations $x_i$ with emission distribution $F()$ are drawn as 
	\begin{align}
\begin{aligned}
		G &\sim \textrm{DP}(\alpha, G_0),\\
		\theta_i \mid G &\sim G,\\
		x_i\mid \theta_i &\sim F(\theta_i).
\end{aligned}
	\end{align}

In order to do exact but parallel MCMC sampling for the DPMM on some $P$ processors, \cite{WilliamsonParallelMCMC2013} proposed the following auxiliary variable representation:
	\begin{align}
\begin{aligned}
G_j &\sim \textrm{DP} (\alpha/P, G_0),\\
\boldsymbol{\phi} &\sim \textrm{Dirichlet}(\alpha/P, \cdots, \alpha/P),\\
\gamma_i\mid \phi &\sim \textrm{Multinomial}(\boldsymbol{\phi}),\\
\theta_i\mid G,\gamma_i &\sim G_{\gamma_i},\\
x_i\mid \theta_i &\sim F(\theta_i).
\end{aligned}
	\label{eq:DPpar}
	\end{align}
The auxiliary variable $\gamma_i$ assigns data point $i$ to processor $\gamma_i$. Theorem 1 of \cite{WilliamsonParallelMCMC2013} proves that for $\boldsymbol{\phi}$ and $G_j$ defined as in Eq.~\eqref{eq:DPpar}, $G:=\sum_j \phi_j G_j \sim \textrm{DP}\left(\sum_j \alpha/P, \frac{\sum_j (\alpha/P) G_0}{\sum_j \alpha/P}\right) = \textrm{DP}(\alpha, G_0)$. Therefore, the marginal distribution for $\theta_i$ and $x_i$ remain the same as in the original DPMM representation. Importantly, conditional on the processor allocations $\boldsymbol{\gamma}$, the data points are distributed as independent DPMMs on $P$ machines, which enables independent sampling of cluster indicators in parallel. In our housing price dynamic model, we leverage this auxiliary variable framework in order to allocate entire data streams to multiple machines. The resulting steps of parallel MCMC sampling of the cluster indicators $z_i$ in our model are described in Supplement \ref{App:parallelDPMM}.

Beyond parallelizing the sampler, we additionally ameliorate the computational burden associated with the likelihood evaluations by deriving a simplified Kalman filter exploiting the specific structure of our model.  In particular, for each data stream we only need two sufficient statistics $\left( \bar{\psi}_{t,i}, L_{t,i} \right)$ instead of all of the house-level transactions, where $\psi_{t,i,l}$ is the adjusted sales price for the $l$th sale in tract $i$ at time $t$ after removing the hedonic effects. The sufficient statistic $ \bar{\psi}_{t,i}$ is the mean of the adjusted individual sales prices and $L_{t,i}$ the number of sales for tract $i$ at time $t$. We can think of the simplified Kalman filter as a filter with observation sequence given by the $p^{(k)}$-dimensional vector of mean sales prices for census tracts in that cluster.  
This algorithm then has complexity $O((p^{(k)})^3T)$.  Although the complexity of the algorithm has not changed (assuming $p^{(k)}<n^{(k)}$), the practical implementation details are simplified leading to significant runtime speedups. We experimented on empirical data that has one cluster of 21 census tracts, with $15,855$ observations over 195 months. We repeat the likelihood evaluation 1000 times. The Kalman filter utilizing the Woodbury identity takes 499 seconds, while the simplified Kalman filter with sufficient statistics only takes 232 seconds, saving more than half of the compute time. This optimized Kalman filtering algorithm for performing likelihood evaluations using sufficient statistics is provided in Supplement \ref{AppSub:condLike_suffStat}.


\section{Model Validation by Simulation}\label{sec8:simulation}

\subsection{Settings}

\begin{figure}[t!]
\centering
\includegraphics[width=1\linewidth]{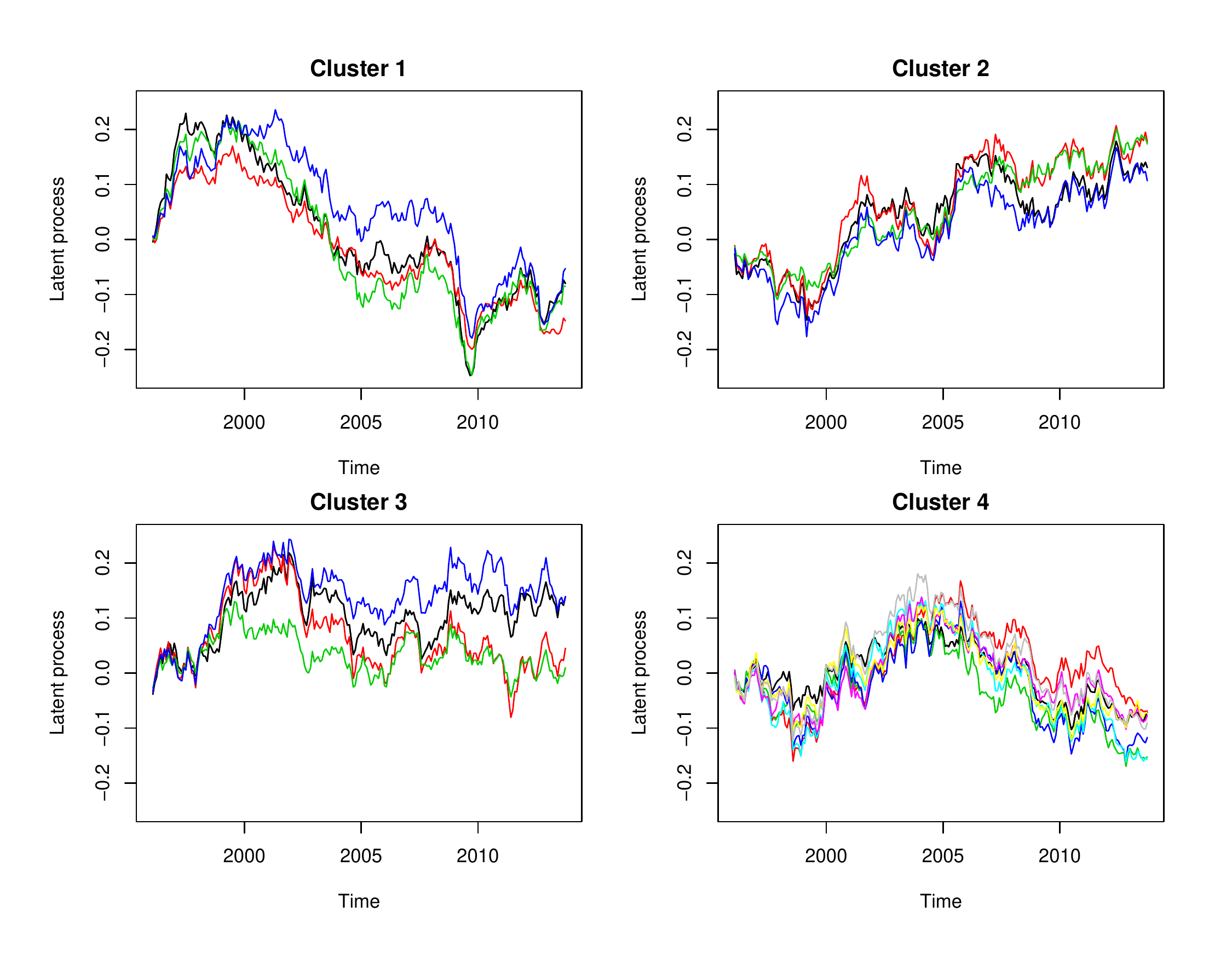}
\caption{Simulated latent price processes for 20 census tracts from 4 clusters. Traces within each plot correspond to specific census tracts in each cluster.}
\label{fig:simuX}
\end{figure}

\begin{figure}[t!]
\centering
\includegraphics[width=1\linewidth]{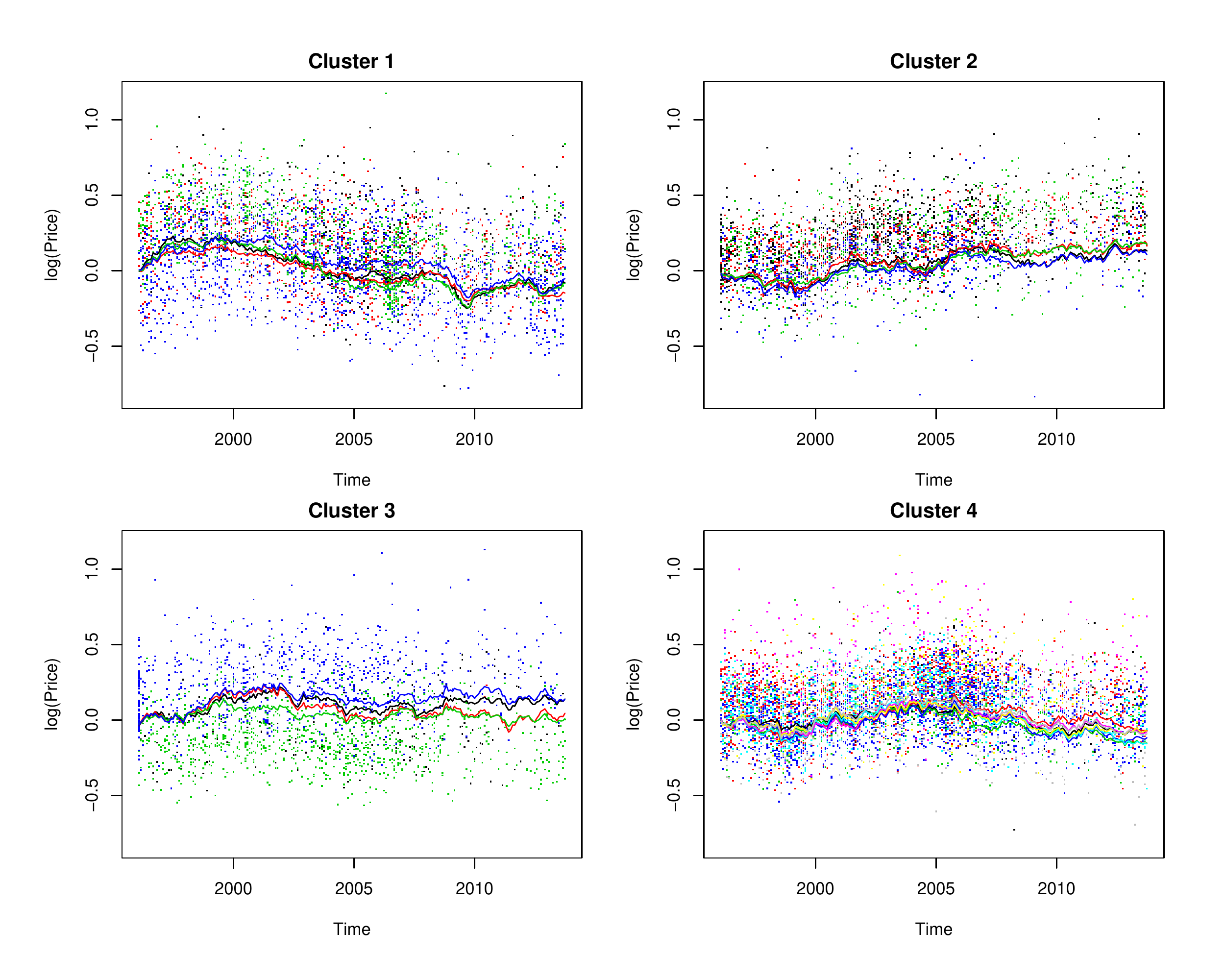}
\caption{Simulated latent process (\emph{solid lines}) and sales prices (\emph{dots}) for the 20 clustered census tracts for each of the 4 ground truth clusters.}
\label{fig:simuY_withBeta}
\end{figure}

We first validate our model using simulated data with aspects set to match our real data analysis of Section~\ref{sec9:realData}. Specifically, we simulated 20 data streams corresponding to sales in 20 census tracts from January 1997 to September 2013, a period of 213 months. The 20 tracts are pre-assigned to four clusters of size 4, 4, 4 and 8 census tracts, respectively. First, we generated latent price processes, $x_{1:T,i}$, for each tract according to Eqs. (\ref{eqn:simpleMod1}) and (\ref{eqn:simpleMod3}) (see Figure~\ref{fig:simuX}). Note that the tracts within each cluster  have similar price dynamics, as intended by our model. Second, we generated the observed sales prices, $y_{t,i,l}$, according to Eq. (\ref{eqn:simpleMod2}). The sales dates and house hedonics are taken from 20 randomly sampled tracts in the City of Seattle, so as to match the real-data frequency of observations and house characteristics.  The resulting generated sales prices are shown in Figure~\ref{fig:simuY_withBeta}. 

\subsection{Results}\label{sec8.2:simulationResults}

\begin{figure}[t!]
\centering
\includegraphics[width=1\linewidth]{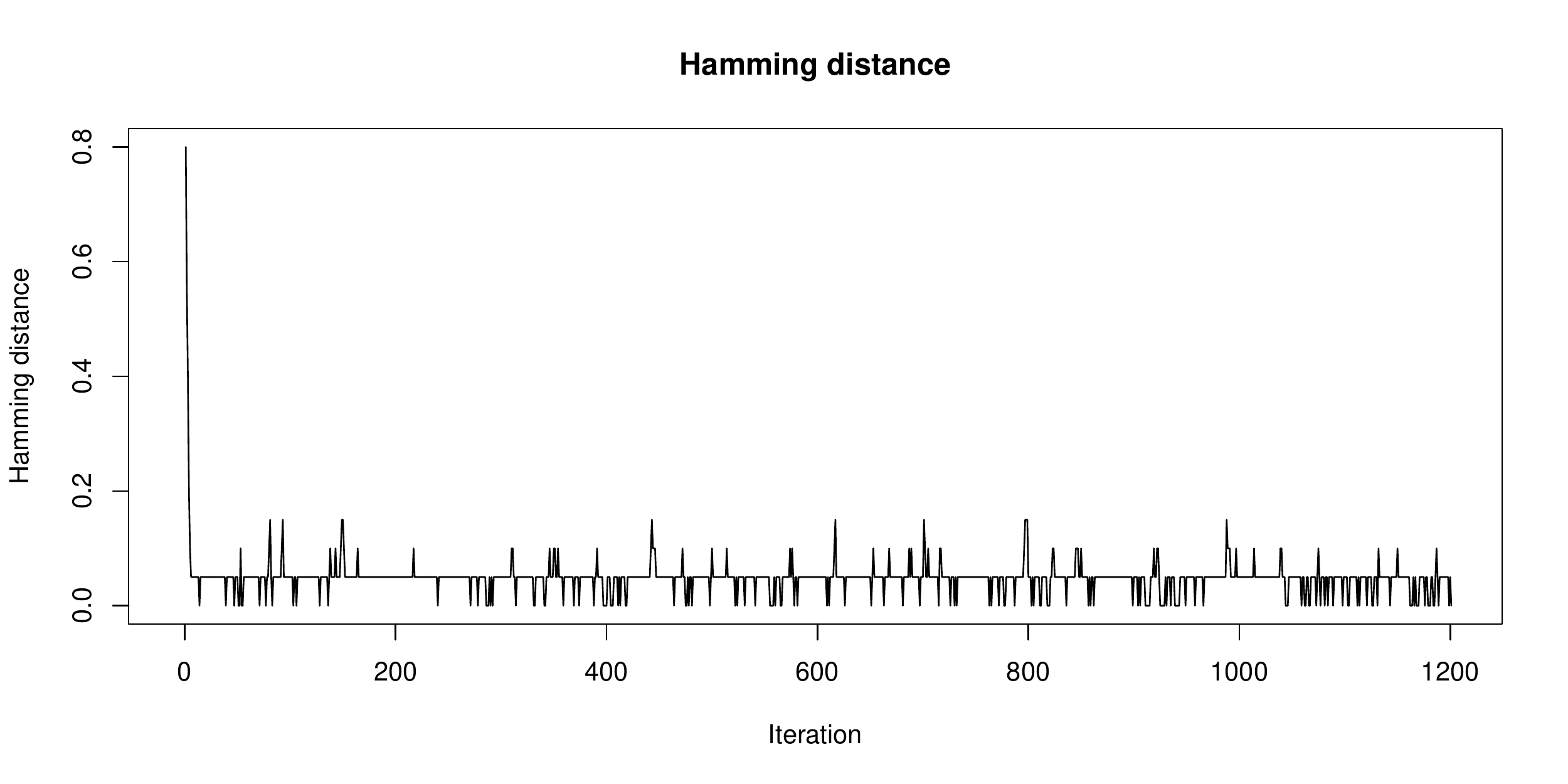}
\caption{Hamming distance between posterior samples of cluster indicators and true cluster memberships (after an optimal mapping) as a function of Gibbs iteration.}
\label{fig:simuHamming}
\end{figure}

We ran the MCMC sampler for 1200 iterations on the simulated data. Figure~\ref{fig:simuHamming} shows the normalized Hamming distance between the estimated and true cluster assignments after an optimal mapping between the sets of labels \citep{munkres1957algorithms}, demonstrating successful recovery of the underlying clusters.  We see that our sampler converges very rapidly.
%

\begin{figure}[t!]
\centering
\includegraphics[width=1\linewidth]{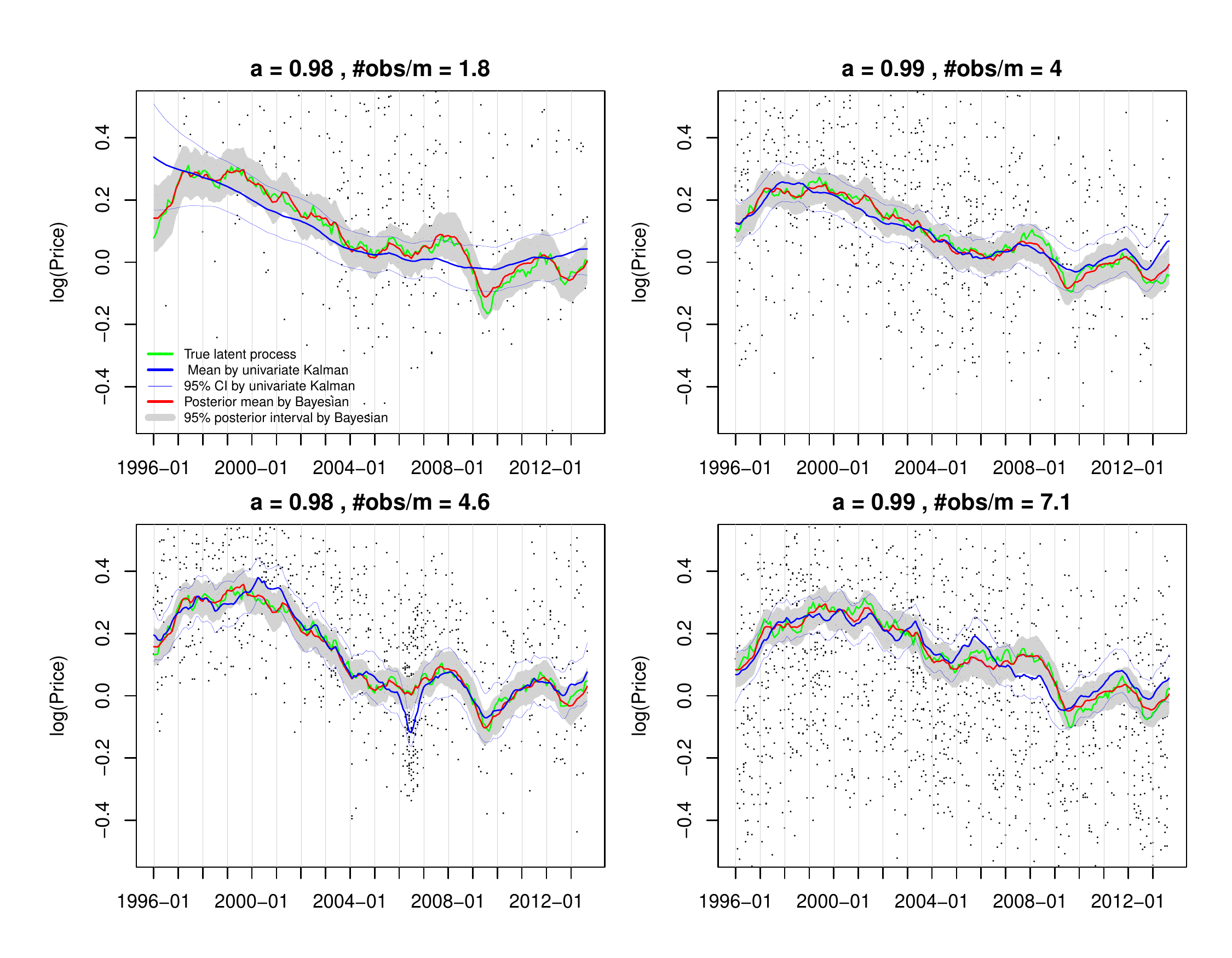}
\caption{Performance of estimating the latent price processes, $x_{t,i}$, shown in green for the 4 census tracts in Cluster 1. The posterior mean and 95\% posterior intervals for our proposed nonparametric clustering-based model are shown in red and shaded gray, respectively.  The blue lines correspond to the independent Kalman smoother baseline approach.}
\label{fig:recoverTrueX}
\end{figure}

Given sparse observations per month at the census tract level, Figure~\ref{fig:recoverTrueX} demonstrates that our posterior estimate of the latent processes nicely tracks the true latent dynamics for each census tract. As a baseline comparison, we considered applying a Kalman smoother independently on each census tract. Unsurprisingly, without sharing observations from similar tracts, the baseline approach fails when the observations are sparse. For other census tracts, please refer to Supplement \ref{App:SimulationResults}.

\begin{figure}[t!]
\centering
\includegraphics[width=1\linewidth]{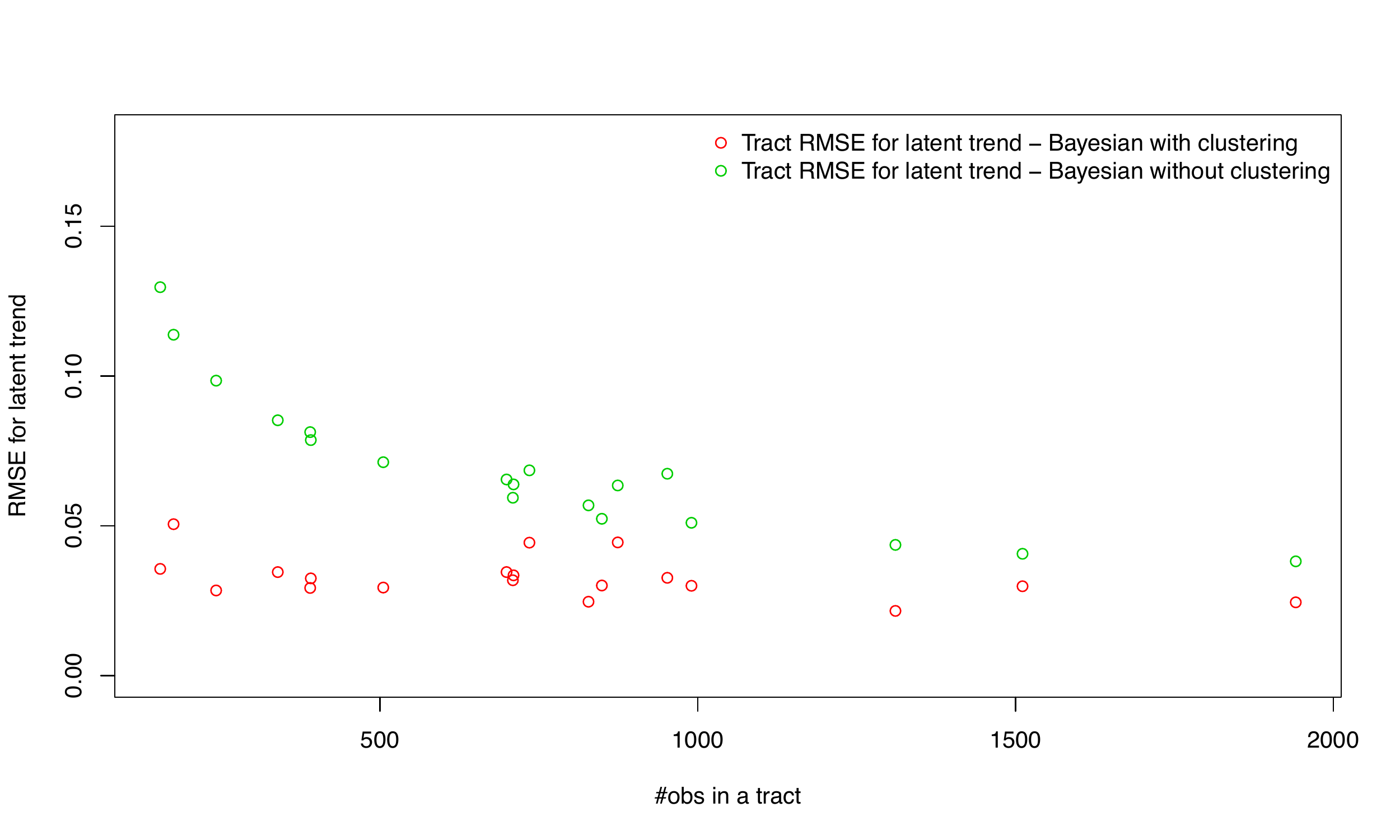}
\caption{Prediction error (RMSE) in latent trend by tract of varying number of observations.}
\label{fig:simuRMSE_x}
\end{figure}

To evaluate the importance of the DP clustering beyond the benefits provided by our hierarchical Bayesian dynamic model, we compare results by enabling / disabling clustering in our proposed model. For the latter, we fixed each census tract to form its own cluster and simply did not resample the cluster indicators in our MCMC. Figure \ref{fig:simuRMSE_x} shows the test set RMSE for predicting the latent trend $\mathbf{x}$ as a function of the number of observations in the census tract. For tracts with fewer observations, the clustering method provides substantial improvement in prediction error.  As expected, when observations are abundant, the improvement diminishes. 

We also experimented with other simulation scenarios, summarized in Table \ref{tab:simuScenarioResults}. When the latent factor processes have relatively large factor loadings (large $\mu_\lambda$) leading to large noise variance on the latent price dynamics, the improvement in predicting latent trends $\mathbf{x}$ are very significant compared to the model without clustering. However, even under such scenarios, the improvement in predicting the observations $y_{i,t,l}$ themselves is not as large since the hedonic effects dominate the observed price.  Importantly, we note that \emph{house level prediction is not our goal}; instead we are interested in the intrinsic price dynamics $\mathbf{x}$ themselves, which form our fine-resolution index. 

\begin{table} [t!]
\caption{Three simulation scenarios and results on out-of-sample prediction of latent trends $x_{1:T,i}$ and house prices $y_{i,t,l}$. We compare our proposed Bayesian model both with and without the DP-based nonparametric clustering component.}
\label{tab:simuScenarioResults}
\begin{tabular}{lrrrr}
\hline
 & & No clustering & Clustering &  Improvement \\
\hline
$\mu_a=0.99$, $\mu_\lambda=0.015$ & RMSE in $x$ & 0.0234 & 0.0191  & 18\% \\
 &  RMSE in $y$ &  0.1192 & 0.1186 &  0.5\% \\
$\mu_a=0.99$, $\mu_\lambda=0.15$ &  RMSE in $x$ & 0.0737 &  0.0335 & 55\%  \\
 &  RMSE in $y$ &  0.3375 & 0.3211 & 4.9\% \\
$\mu_a=0.60$, $\mu_\lambda=0.15$ &  RMSE in $x$ & 0.0786 & 0.0313 & 60\% \\
 &  RMSE in $y$ &  0.1335 & 0.1219 &  8.7\% 
\\ \hline
\end{tabular}
\end{table}

\section{Housing Data Analysis}\label{sec9:realData}

\begin{figure}[t!]
\centering
\includegraphics[width=0.6\linewidth]{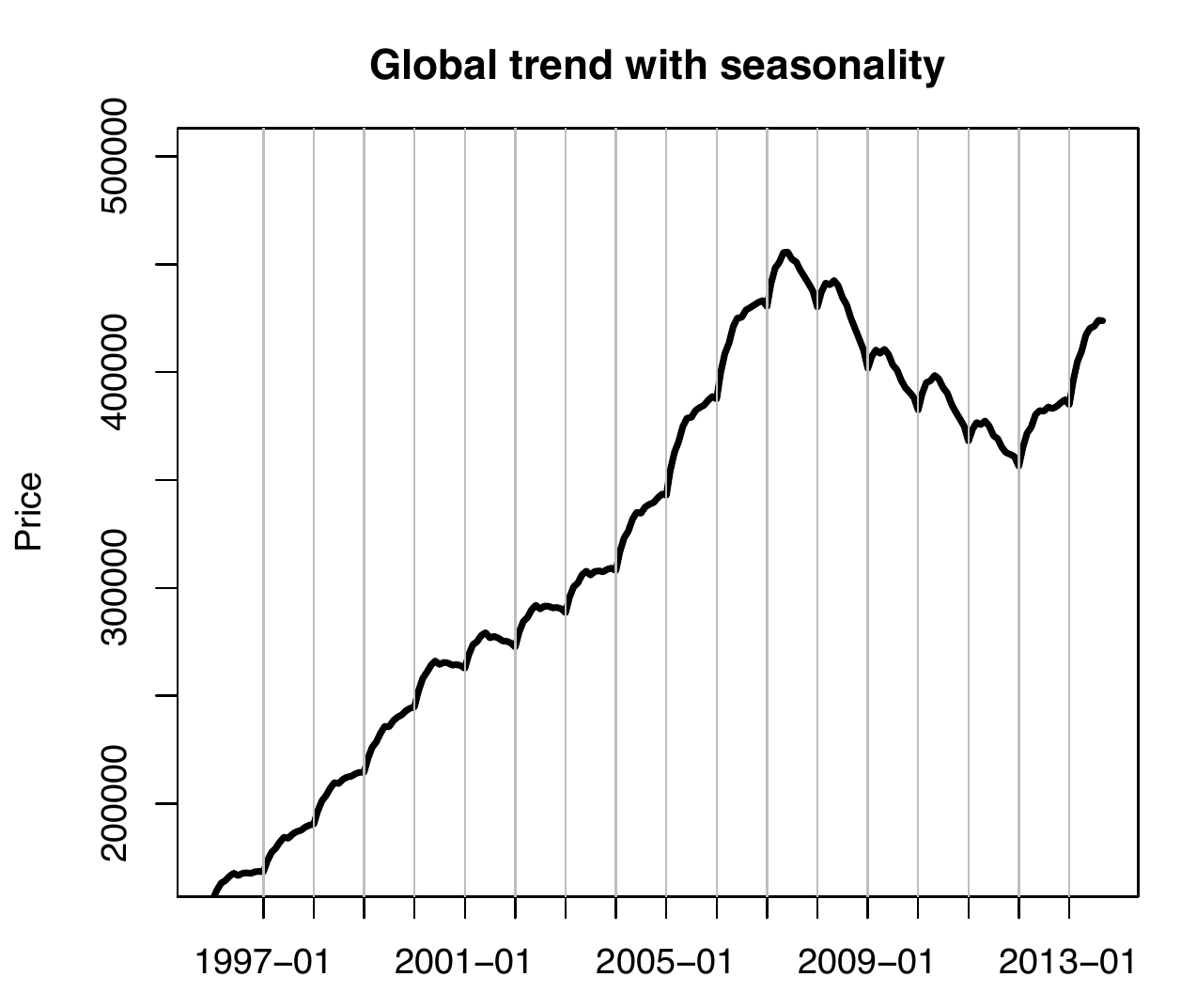}
\caption{Estimated global trend using the seasonality decomposition approach of \cite{seasonalDecomposition1990}, after adjusting for hedonic effects.}
\label{fig:GlobalTrend}
\end{figure}

We now turn to our housing data analysis based on the City of Seattle data described in Section~\ref{sec2:data}.  Recall our goal of forming a census-tract level index.  For simplicity, we have assumed a separately estimated global trend ($g_t$ in Eq.~\eqref{eqn:fullMod1}), which captures the city-wide price dynamics, though it would be straightforward to incorporate joint inference of $g_t$ in our MCMC.  For our experiments in this section, we base this global trend on an estimate formed as follows.  We first consider a non-tract-specific regression akin to Eq.~\eqref{eqn:simpleMod2} in order to remove the hedonic effects: $y_{t,i,l} =  \alpha_0 + \sum_t\alpha_t I(t) + \sum_{h=1}^H \beta_{h} U_{l,h}+v_{t,l}$, where $\alpha_t$ captures the monthly effect and $\beta_h$ the hedonic effects on the global trend.  The noise $v_{t,l}$ is independent across time and sales.  Note that in aggregate, we have roughly 640 observations per month on average.  After removing the hedonic effects, we then apply the seasonal decomposition approach of \cite{seasonalDecomposition1990} to decompose the estimated global trend into a trend component, seasonal component, and noise; we discard the noise term.  The resulting global trend (Figure~\ref{fig:GlobalTrend}) has a small but significant seasonal effect.  This can be mostly attributed to the changing supply of houses during the year: very few homes are listed in November and December so that transactions that occur in that period are leftover inventory or have other special circumstances.

To asses our model, we randomly split the sales \emph{per census tract} into a 75\% training and 25\% test sets.  On the training set, we ran three MCMC chains for $15,000$ iterations from different initial values, discarding the first half as burn-in and thinning the remaining samples by 5. We used the scale reduction factor of \cite{GerlmanRubin} to check for convergence of chains.

\begin{figure}[t!]
\centering
\includegraphics[width=0.5\linewidth]{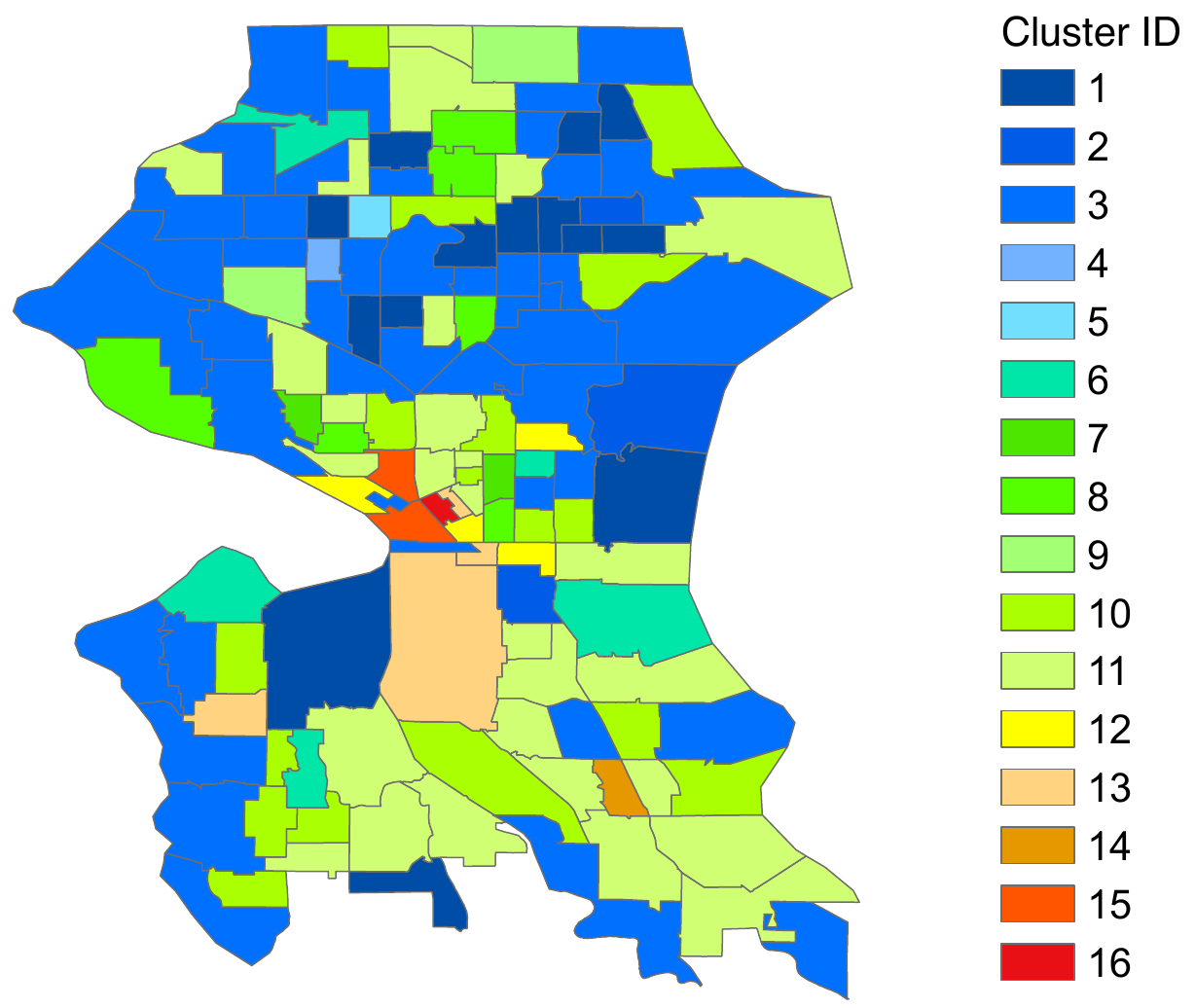}
\caption{Map of clusters under the MAP sample.  The cluster labels and associated map colors are selected to indicate the level of deviance of the cluster's average (across tracts) latent trend from the global trend. Blue (1) represents a small deviance while red (16) represents the largest.}
\label{fig:ClusterMap}
\end{figure}

\begin{figure}[t!]
\centering
\includegraphics[width=1\linewidth]{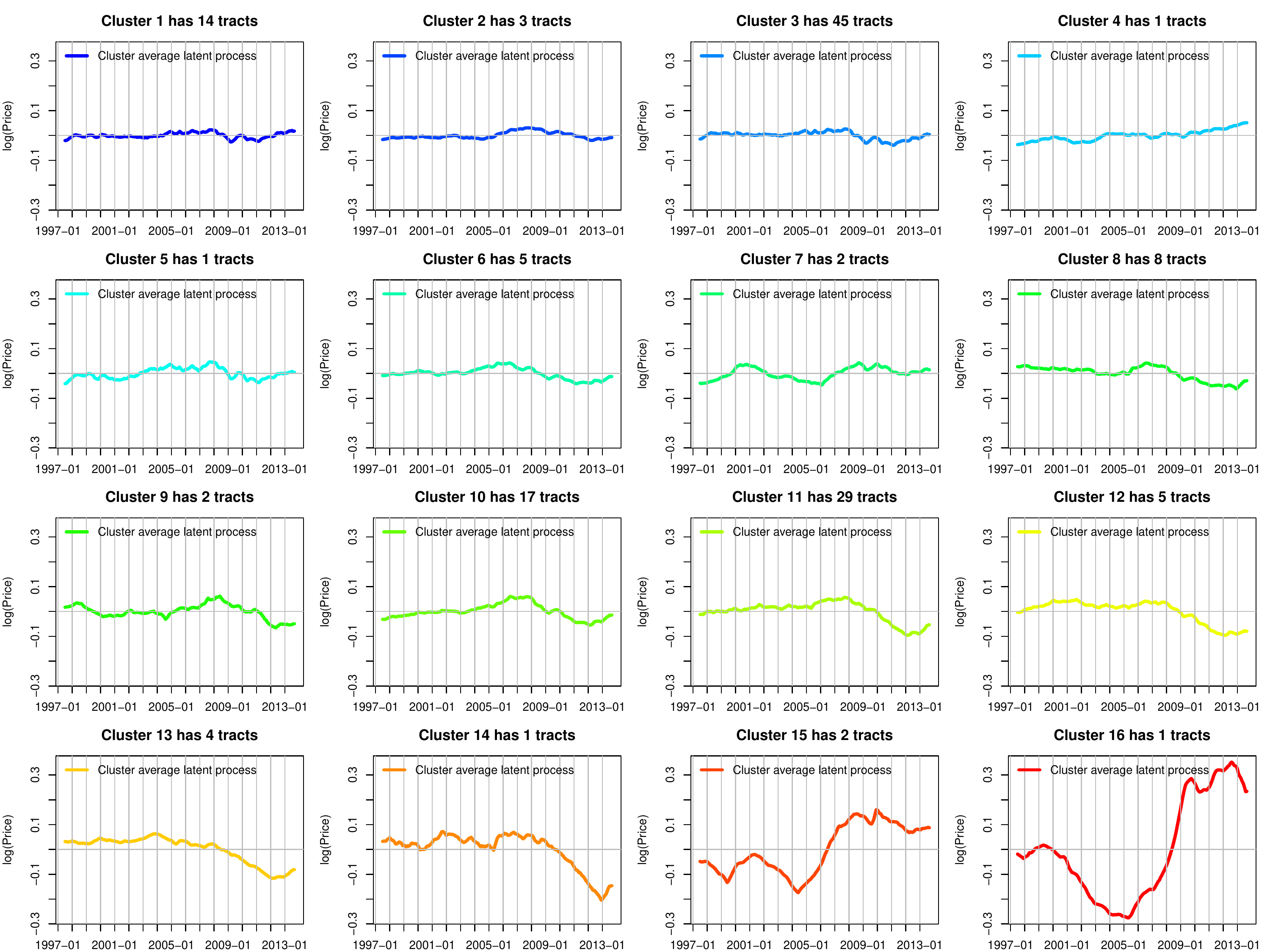}
\caption{Under the MAP sample, cluster-average intrinsic price dynamics computed by averaging $\mathbf{x}_{1:T,i}$ over all $i$ with $z_i=k$ for $k=1,\dots,16$. The color scheme is the same as in Figure~\ref{fig:ClusterMap}.}
\label{fig:ClusterPanel}
\end{figure}

Figure~\ref{fig:ClusterMap} provides an illustration of the resulting 16 census tract clusters associated with the maximum a posteriori (MAP) sample (i.e., the sample with largest joint probability).  The log intrinsic price dynamics associated with each of these clusters, averaged over census tracts assigned to the cluster, are shown in Figure~\ref{fig:ClusterPanel}.  Cluster 15 and 16 have the most dramatic trend.  They include census tracts from the downtown Seattle area where the houses are almost exclusively condos and have unique supply and demand dynamics. Cluster 11 and 13 are mostly low-income areas with less expensive housing where the housing recovery has been slower. The biggest difference between the clusters occurs during the 2006-2012 time period which spanned the housing boom followed by the bust. Intuitively, different regions were affected differently by this highly volatile period. Supplement \ref{App:realDataResults} shows the cluster average index in raw price scale.

For this MAP sample, the University District (U-District) census tract highlighted in Section~\ref{sec1:introduction} gets assigned to Cluster 3---the largest cluster---driven by ``the rich get richer'' property of the CRP prior. However, when examining all collected posterior samples, 57\% of the time the U-District does not share a cluster with \emph{any} of its neighbors and 86\% of the time it does not share a cluster with more than one neighbor. The lack of a hard-coded spatial structure in our model is what enables such heterogeneous spatial effects to appear; instead, our DP-based cluster model allows for a flexible dependence structure by discovering regions with similar price dynamic patterns. 

\subsection{Comparison with other methods} \label{sec9.1:SeattleResultsComparison}
We compared our Bayesian nonparametric approach with the Case-Shiller housing index \citep{CaseShiller1987} described in Section~\ref{sec1:introduction}.  Even though our goal is not house-level prediction, it is one metric by which we can assess our fit.  Since the Case-Shiller method is based on repeat sales only and does not include hedonics, it is not well-suited to predicting house-level prices.  In order to fairly compare our approach with Case-Shiller, we treated the Case-Shiller index as the latent process $\mathbf{x}$ in our model, and then fit a regression model with tract-specific hedonic effects as in Eq.~\eqref{eqn:simpleMod2}. The estimated hedonic effects together with Case-Shiller index are then used to predict the house prices. Due to the scarcity of repeat sales observations localized at tract level, the Case-Shiller index can only be computed at 8 of the 140 tracts.  To maintain a tract-level comparison, if the Case-Shiller index is not available for a given tract, we continue up the spatial hierarchy examining zip code and city levels until there is a computable index that can serve as $x_{t,i}$ in our prediction.  That is, we use the finest resolution Case-Shiller index available at any house location to predict house prices. In Table~\ref{tab:numObsAvailableUnderDiffMethods}, we summarize the number of house-level predictions that are based on the Case-Shiller city, zip code, or tract level indices; we also include the number of tracts for which our analyses relied on city and zip code levels, or were able to use tract-level indices directly.


\begin{table}[t!]
\caption{For our predictive performance comparison summarized in Table~\ref{tab:resultsComparison}, the number of tracts and individual houses (in test set) that rely on using city, zip code, or tract-level indices with the Case-Shiller method.  Our Bayesian method always uses a tract-level index.}
\label{tab:numObsAvailableUnderDiffMethods}
\begin{tabular}{lrrrr}
\hline
& Case-Shiller  & Case-Shiller  & Case-Shiller  & Bayesian \\
& City & Zip Code & Census Tract & Census Tract\\
\hline
\# tracts using & 11 & 121 & 8 & 140\\
\# observations using & 1,294 & 26,576 & 3,248 & 31,118\\
\hline
\end{tabular}
\end{table}

Our Bayesian model can successfully produce value indices for all tracts. To predict house-level prices, we use the posterior predictive distribution approximated by our MCMC posterior samples:
\begin{eqnarray}
P(y_{t,i}^*|{\mathbf Y}) = \int_{\theta} P(y_{t,i}^*|\theta)P(\theta | {\mathbf Y}) d\theta \approx \sum_{m=1}^M p(y_{t,i}^*\mid \theta^{(m)}),
\end{eqnarray}\label{eqn:predictiveYdist}
where $y^*$ is the new data point, ${\mathbf Y}$ denotes the training data and $\theta$ represents parameters with $\theta^{(m)}$ the $m$th MCMC sample.  Since $p(y_{t,i}^*\mid \theta^{(m)})$ does not have an analytic form, we simulate a set of $y_{t,i}^*$ for each $\theta^{(m)}$ using Eq.~\eqref{eqn:simpleMod2}. We then use the mean of these posterior predictive samples as prediction for any house in the test set. 

For all of our comparisons, we used the same training and test split.  In Table~\ref{tab:resultsComparison}, we summarize the out-of-sample predictive performance with five metrics: root mean squared error (RMSE) in price, mean / median / 90\% quantile of absolute percentage error (Mean APE, Median APE, 90th APE), and the popular industry metric of proportion of house sales within 10\% error (P10). Importantly, we highlight again that house sales predictions are largely hedonics driven.  Since we constructed all methods using the same hedonics model, we do not expect to see large differences in numbers.  Regardless, we see notable improvements using our proposed index, with uniformly better predictive performance as compared to the Case-Shiller index at the finest resolution available.  Over all houses in the test set, our method has an 11.2\% improvement in RMSE and about 5\% improvement in other metrics. 

We then break the analysis down by deviation of the inferred latent trend from the global trend.  For the top 5\% tracts with most dramatic local price dynamics (measured in L2 distance of posterior mean latent trend over time), we see even more dramatic improvements over Case-Shiller: a 15.5\% decrease in RMSE and 21.7\% in 90th percentile APE. The latter measure indicates a significant reduction in the tail of the error distribution. That is, not only are we better able to capture these more volatile tracts, we are also having the most dramatic improvements on the hardest-to-predict houses.  These effects can be explained as follows. By not hard-coding a neighborhood structure, we see in Figure~\ref{fig:Map_latentTrend} that certain regions (e.g. the U-District) do not get shrunk to trends in neighboring tracts. At the same time, our hierarchical Bayesian model with clustering still enables sharing of information to improve estimates, as we see in Table \ref{tab:resultsComparison}. It is not surprising to see the most significant improvements being for the most highly volatile tracts: these are the tracts for which providing a robust fine-scale index is so important in order to capture the deviation from the global trend.  


\begin{table} [t!]
\caption{Predictive performance comparison of index methods using various measures: root mean squared error (RMSE), mean absolute percentage error (Mean APE), median absolute percentage error (Median APE), 90th percentile absolute percentage error (90th APE) and proportion within 10\% error (P10).}
\label{tab:resultsComparison}
\begin{tabular}{lrrrr}
\hline
 & Case-Shiller index & Bayesian index & \\
 & at finest resolution &  at census tract & {\it Improvement} \\
 & w/ tract hedonic effects & level &  \\
\hline
\multicolumn{4}{l}{\it All observations in test set (31,118 data points)} \\
\qquad RMSE & 137,600 & 122,139 & 11.2\% \\
\qquad Mean APE & 0.1734 & 0.1636 & 5.6\% \\
\qquad Median APE & 0.1294 & 0.1236 & 4.5\% \\
\qquad 90th APE & 0.3607 & 0.3427 & 5.0\% \\
\qquad P10 & 0.3985 & 0.4190 & 5.1\%\\
\\
\multicolumn{4}{l}{\it Top 5\% tracts with most dramatic latent trends (1,111 data points)} \\
\qquad RMSE & 91,627 & 77,399 & 15.5\% \\
\qquad Mean APE & 0.2045 & 0.1748 & 14.5\% \\
\qquad Median APE & 0.1403 & 0.1259 & 10.3\%\\
\qquad 90th APE & 0.4699 & 0.3679 & 21.7\% \\
\qquad P10 & 0.3816 & 0.4113 & 7.8\% \\
 \hline
\end{tabular}
\end{table}


Table \ref{tab:CSIcity_CSIzip_Bayes_numObsCategories} lists the improvement in predictive performance of our Bayesian tract index over using the Case-Shiller index computed at a city or zip code level. The most significant improvement is for houses in tracts with fewer sales (lower 5\% tracts). For example, we see a 16\% improvement in 90th percentile APE for these data-scarce tracts, for which the tail of the error distribution is important and hard to characterize. We might expect that our method provides less improvement over the Case-Shiller index at the zip code than city level. Interestingly, as the spatial resolution goes finer from city to zip code level, the Case-Shiller index suffers from worse predictive performance in most cases. This result validates that this popular index method is ill-suited to the task of constructing a housing index for small regions where transactions are scarce.  

\begin{table} [t!]
\caption{Predictive performance improvement of our Bayesian tract index over Case-Shiller City and Zip code indices for tracts of different sales frequency, using various measures: mean absolute percentage error (Mean APE) and 90th percentile absolute percentage error (90th APE).}
\label{tab:CSIcity_CSIzip_Bayes_numObsCategories}
\begin{tabular}{lrr}
\hline
 & Improvement over & Improvement over \\
 &  Case-Shiller City index &  Case-Shiller Zip Code index \\
\hline
\multicolumn{3}{l}{\it Top 5\% tracts with most sales (3,569 data points)} \\
\qquad Mean APE & 3.1\% & 4.8\% \\
\qquad 90th APE & 1.2\% & 2.9\% \\
\\
\multicolumn{3}{l}{\it Middle 50\% tracts (14,507 data points)} \\
\qquad Mean APE & 4.6\% & 7.2\% \\
\qquad 90th APE & 5.1\% & 7.1\% \\
\\
\multicolumn{3}{l}{\it Lower 5\% tracts with least sales (188 data points)} \\
\qquad Mean APE & 8.5\% & 5.4\% \\
\qquad 90th APE & 15.5\% & 16.0\% \\
\hline
\end{tabular}
\end{table}

We now examine the impact of our various modeling components on our overall performance. We start by comparing the performance of our approach with simpler dynamical models.  In particular, we compare against models that treat each census tract independently.  Both use the per-tract dynamics specified in Eqs.~\eqref{eqn:simpleMod1}-\eqref{eqn:simpleMod2}, though one of our comparisons omits the hedonics term.  In both cases, the intrinsic price dynamics and associated model parameters are inferred independently for each census tract using a Kalman smoother embedded in an expectation maximization (EM) procedure.  The results are summarized in Table~\ref{tab:resultsComparison_breakdown_2}. (Note that the last column of Table~\ref{tab:resultsComparison_breakdown_2} coincides with that of Table~\ref{tab:resultsComparison}, and is repeated for readability.)  We see reduced predictive performance at each stage of breaking down our Bayesian dynamical model.  Additionally, as motivated by the results of Table~\ref{tab:simuScenarioResults}, we would expect even larger improvements in the estimation of the target index $\mathbf{x}$, though such an evaluation is not feasible here since we do not have the true index value.

\begin{table} [t!]
\caption{Predictive performance comparison on variants of the proposed Bayesian nonparametric model using the same metrics as in Table~\ref{tab:resultsComparison}.}
\label{tab:resultsComparison_breakdown_2}
\begin{tabular}{lrrr}
\hline
& Univariate  & Univariate & Bayesian \\
&  Kalman Smoother & Kalman Smoother & clustering \\
& w/o hedonics & w/ hedonics & \\
\hline
RMSE  & 262,075 & 194,562  & 122,139\\
Mean APE & 0.3698 & 0.2746 &  0.1636\\
Median APE & 0.2854 & 0.2238  &   0.1236\\
90th APE & 0.7634 & 0.5584 &  0.3427\\
P10 & 0.1907 & 0.2346   &  0.4190
\\ \hline
\end{tabular}
\end{table}

We now turn to the central focus of the paper and assess the quality of the index itself.  Since there is no ground truth or direct performance metric, we use the Zillow Home Value Index (ZHVI) as a proxy.  As mentioned in Section~\ref{sec1:introduction}, the index is formed by taking the median of Zillow house-level estimates of value and provides a stable empirical estimate at fine-scale regions.  In addition to comparing our Bayesian index to Case-Shiller, we also consider a model in which the the DP-based clustering is removed, treating each census tract as its own cluster. This model still represents a hierarchical Bayesian dynamic model.  Since the Case-Shiller method is not computable for most of the census tracts, we focus our analysis at the zip code level. For the Bayesian index with or without DP-based nonparametric clustering, the zip code index is constructed by averaging the census tract indices within the a zip code.  

\begin{figure}[t!]
\centering
\includegraphics[width=0.8\linewidth]{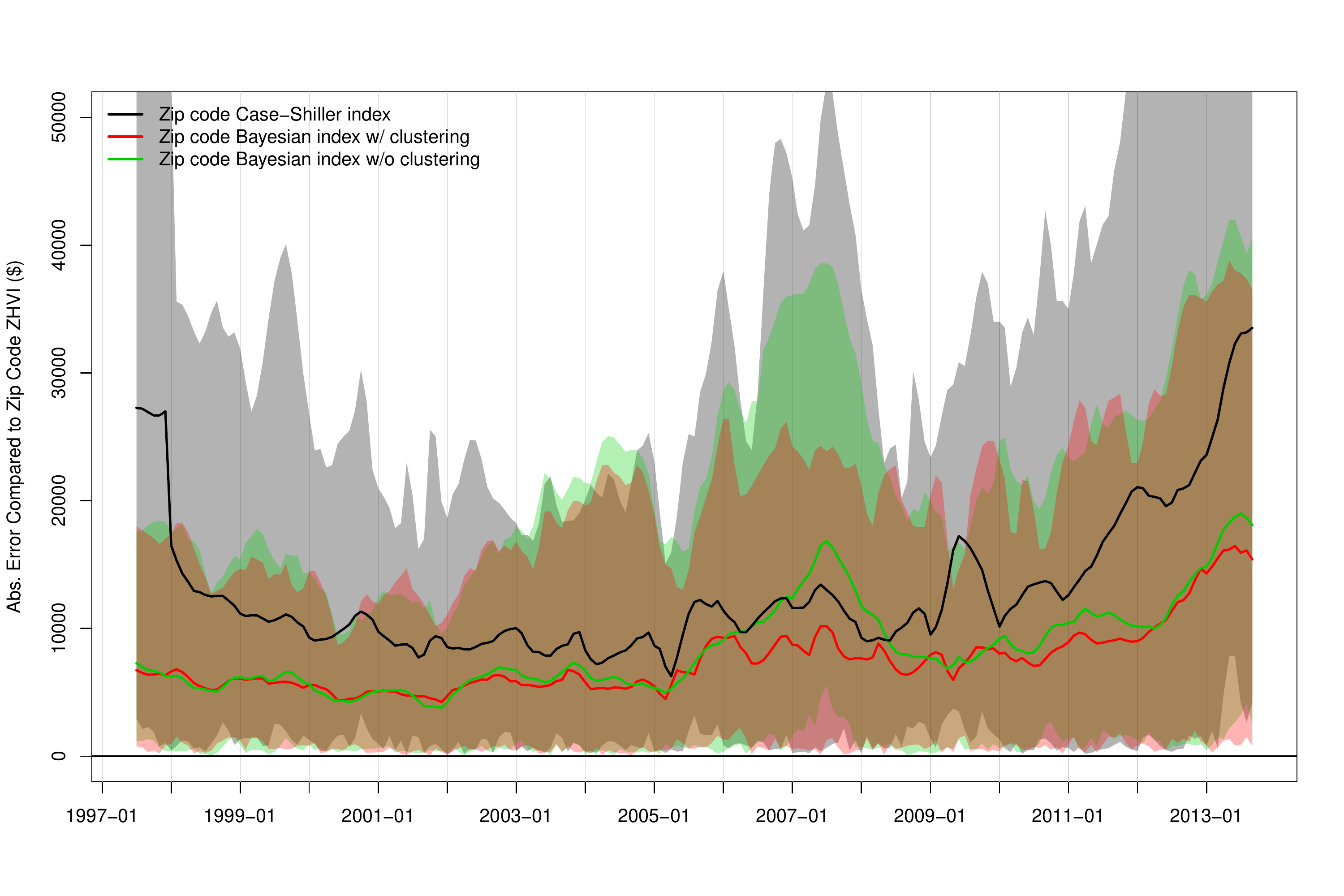}
\caption{Treating the Zillow Home Value Index (ZHVI) as a surrogate ground truth, errors of various index methods relative to ZHVI at the zip code level. Examining performance across zip codes, the mean absolute error (\emph{red line}) and 90\% interval (\emph{shaded red}) of our proposed Bayesian index is compared to that of the Bayesian index without the DP-based nonparametric clustering component (\emph{green} and \emph{shaded green}) and the Case-Shiller zip code index (\emph{black} and \emph{shaded gray}). The performance of the Bayesian methods are based on posterior mean estimates.}
\label{fig:compareZipIndex_threeMethods}
\end{figure}

Figure~\ref{fig:compareZipIndex_threeMethods} shows that the Bayesian index with (\emph{red line}) and without (\emph{green line}) the DP-based nonparametric clustering component have significantly different performance during the 2006-2007 period, and to a lesser extent in 2010-2011.  In 2006-2007, the Seattle housing market was in a boom period with high sales and volatility  (see Figure~\ref{fig:compareIndex_salesVolume} in Supplement \ref{App:salesVolume}). After the bust, the housing market started to stabilize in 2010-2011. The market boom and subsequent stabilization were manifested in the different housing sectors in disparate ways. The DP-based clustering, especially in the highly volatile year of 2007, is more closely aligned with the ZHVI, since it is better able to capture the dynamics of the change in value for different housing sectors.  This is because the non-clustering Bayesian hierarchical model shrinks the census tracts with few observations towards a global mean, whereas our clustering model allows atypical census tracts to be shrunk towards a more informed structure, such as the one shown in Figure~\ref{fig:ClusterPanel}.

Figure~\ref{fig:compareZipIndex_threeMethods} also compares the zip code Case-Shiller index (\emph{black line}), 
which is significantly more different from the ZHVI than the proposed Bayesian index (\emph{red line}) during all times. 
Without any kind of sharing information and shrinkage across different regions, the Case-Shiller index has the widest interval among the three methods. The beginning and the end of the study periods are extremely challenging for Case-Shiller index, because of having fewer repeated sales available at the boundary of the study period. In the middle of the series, the difference between Case-Shiller and the ZHVI is especially large during the highly volatile period of 2007. 

\begin{figure}[t!]
\centering
\begin{tabular}{@{\hspace{0mm}}c@{\hspace{1mm}}c@{\hspace{1mm}}c@{\hspace{1mm}}c@{\hspace{1mm}}c@{\hspace{1mm}}c@{\hspace{1mm}}c}
\begin{tabular}{@{\hspace{0mm}}c@{\hspace{1mm}}c@{\hspace{1mm}}c@{\hspace{1mm}}c@{\hspace{1mm}}c@{\hspace{1mm}}c@{\hspace{1mm}}c}
\includegraphics[width=0.49\linewidth]{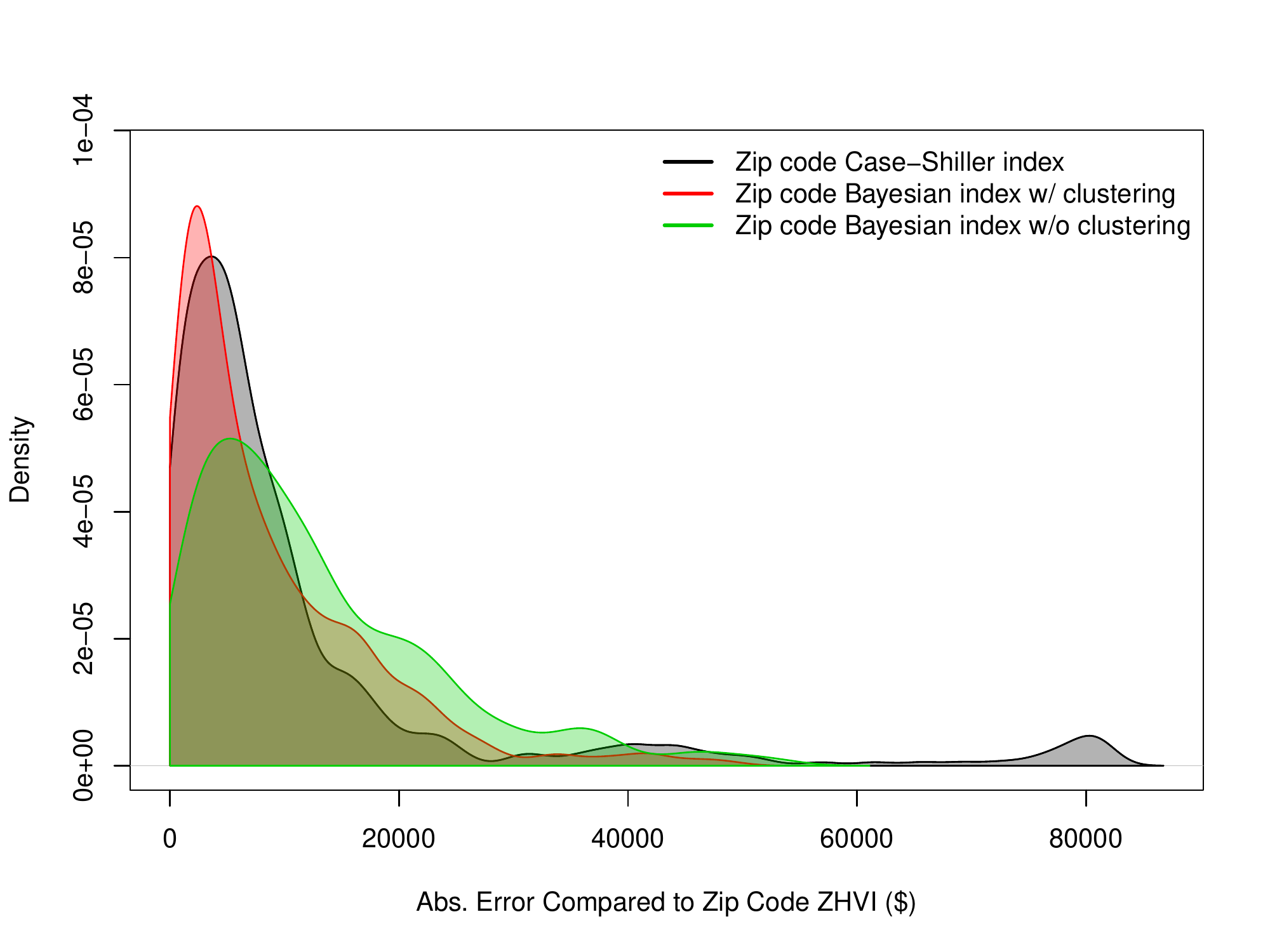}&
\includegraphics[width=0.49\linewidth]{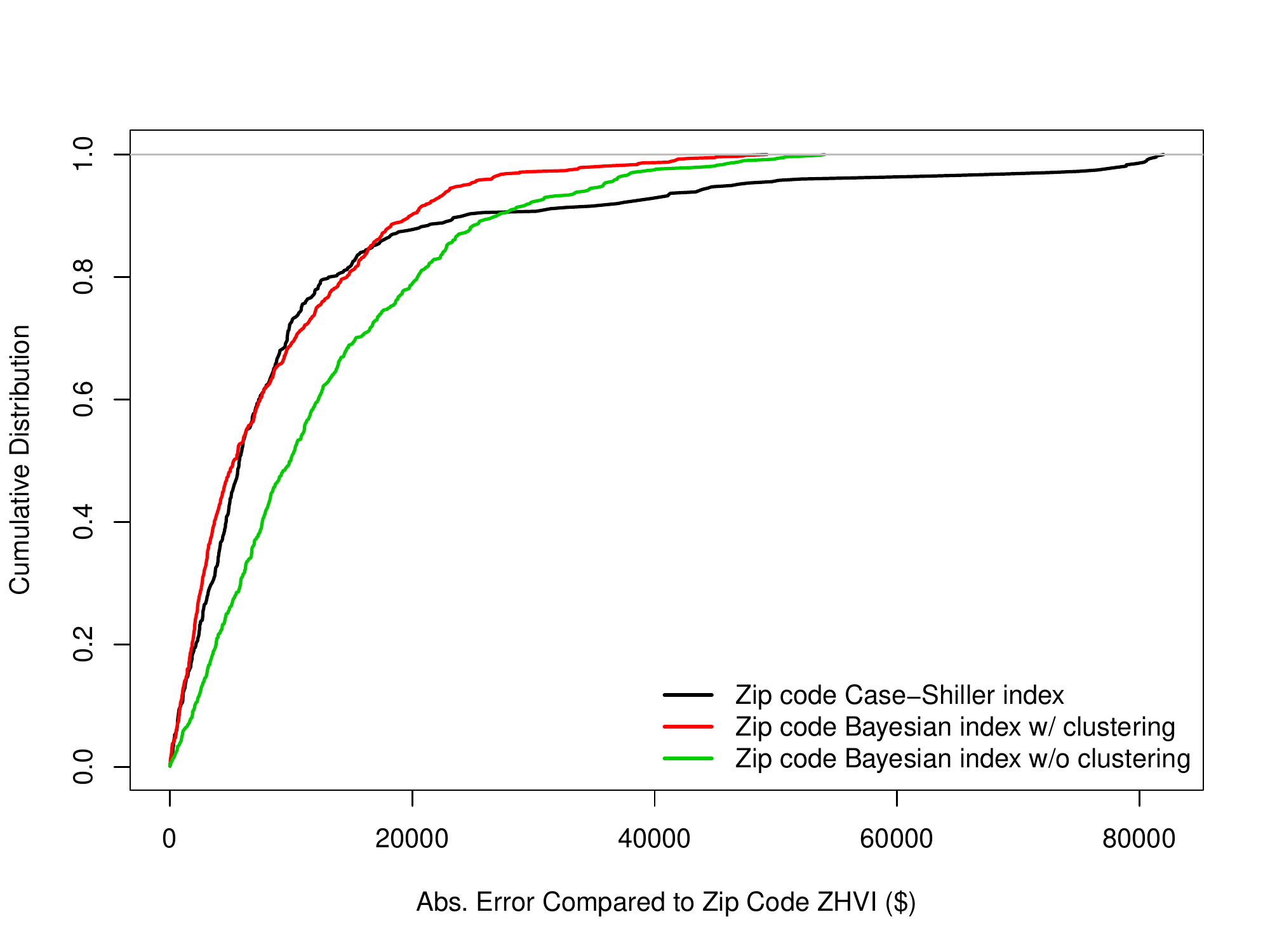}\\
(a) & (b)
\end{tabular}\\
\end{tabular}
\caption{A more detailed examination of the distribution of errors in Figure~\ref{fig:compareZipIndex_threeMethods} during 2007. (a) Estimated density and (b) associated cumulative distribution of the absolute error.  }
\label{fig:compareZipIndex_2007}
\end{figure}

Figure~\ref{fig:compareZipIndex_2007} shows that Case-Shiller (\emph{black}) has a long-tailed distribution of absolute error relative to ZHVI in contrast to the shrinkage provided by the other two Bayesian methods, with the clustering approach clearly the best.  In particular, looking the cumulative distribution of Figure~\ref{fig:compareZipIndex_2007}(b), we see that the Bayesian model without clustering has a lighter tail than Case-Shiller, improving these outlying estimates via shrinkage induced by the hierarchical Bayesian model; however, the Bayesian non-clustering model also has fewer low-error zip codes relative to the Case-Shiller baseline.  In contrast, our proposed Bayesian nonparametric clustering index has as many low-error zip codes as Case-Shiller, tracking this baseline index in the low-error range, but also has many fewer high-error zip codes than either of the comparison methods.  Thus, we see the importance not only of a hierarchical Bayesian approach, but one that leverages structured relationships between regions.

\section{Discussion} \label{sec10:discussion}
We presented a method for constructing a housing index at fine-scale geographical units, with better space-time adjustment and specificity than existing approaches.  In particular, the extreme sparsity of transactions at a fine spatiotemporal granularity poses a significant modeling challenge.  Our proposed dynamical model utilizes a Bayesian nonparametric approach for flexible structure learning to correlate regions that share similar underlying price dynamics.  This model leverages information from the region-specific time series within a cluster, providing a form of multiple shrinkage of individual trend estimates for each region.

Our clustering-based dynamical model avoids a reliance on repeated sales, providing an ability to track price changes in local housing markets. In contrast, constrained by few observations of multiple sales for the same house, classic repeat sales methods are usually only robustly estimated over larger regions, such as zip code or city, which may lack spacial specificity. 

Although sole reliance on repeated sales can be problematic for the reasons described above, one could imagine incorporating a similar idea within our model via a longitudinal trend for the same house in the model. Other extensions include considering longer memory processes with a higher order autoregressive model for the latent trend. We could also add side information, such as crime rate, road network information, and school district ratings, to better inform the clusters of local areas.  Finally, one could consider a pre-specified geographic model combined with our cluster-induced heterogeneous spatial structure as a model of the residuals.

\section*{Acknowledgements}
We would like to thank Stan Humphries, Yeng Bun, Bill Constantine, Dong Xiang and Chunyi Wang at Zillow for helpful discussions and guidance on the data.  Y. Ren was funded in part by a graduate fellowship provided by Zillow.  This work was also supported in part by NSF CAREER Award IIS-1350133, the TerraSwarm Research Center sponsored by MARCO and DARPA, and DARPA Grant FA9550-12-1-0406 negotiated by AFOSR.

\bibliographystyle{Chicago}
\bibliography{houseBib}

\clearpage
\appendix
 \section{Conditional Likelihood of Data in Cluster $k$}\label{App:sampleZ_condLike}
In this section, we describe how to compute the likelihood of the data from all time series assigned to a given cluster $k$, conditioned on the model parameters.  We consider two mathematically equivalent methods: one based on the collection of observations directly, and the other using sufficient statistics of the observed house sales.  In what follows, we drop the cluster index $k$ for simplicity of notation.
 \subsection{Naive Kalman filtering}\label{AppSub:condLike_allObs}
We consider a straightforward extension of the standard Kalman filter recursions to compute the marginal likelihood of all observations in cluster $k$ when there can be multiple observations per time step.  The derivation is as follows. The cluster marginal likelihood can be calculated as
\begin{eqnarray}
\log P({\bf y}_{1:T}) = \sum_{t=1}^T \log P({\bf y}_t|{\bf y}_{1:t-1})
\end{eqnarray}
where the distribution of new observations at time $t$ conditional on past time series is
\begin{eqnarray}
{\bf y}_t|{\bf y}_{1:t-1} \sim \mathcal{N} \left({\bf y}_t \left|C_t\mu_{t|t-1}+D_tU_t, S_t \right. \right). \label{eqnApp: jointYtLikelihood}
\end{eqnarray}
The quantities $\mu_{t|t-1}$ and $S_t$ are obtained by the Kalman filter:
\begin{align}
	\begin{aligned}
\textrm{Predict} \quad \mu_{t|t-1} & =  A \mu_{t-1|t-1}\\
V_{t|t-1} & =  AV_{t-1|t-1}A^T + Q\\
\textrm{Calculate} \quad S_t & =  C_t V_{t|t-1} C_t^T+ R_t\\
\textrm{Kalman gain matrix} \quad K_t &= V_{t|t-1} C_t^T S_t^{-1}\\
\textrm{Filter} \quad \mu_{t|t} &= \mu_{t|t-1} + K_t \left( {\bf y}_t - C_t \mu_{t|t-1}-D_tU_t \right)\\
V_{t|t} &= \left( {\bf I} - K_t C_t \right) V_{t|t-1} 
\end{aligned}
\end{align}
The coefficient matrix $C_t$ is an indicator matrix mapping each observation to its specific census tract. The matrix $D_t$ is a coefficient matrix for hedonic effects.
The filter should be applied to data for all tracts in cluster $k$ together. \\

The purpose of doing filtering here is to evaluate the conditional likelihood of tract $i$ belonging to cluster $k$, given observations of all the other tracts in cluster $k$. The conditional likelihood is
\begin{eqnarray}
P({\bf y}_{1:T, i} | {\bf y}_{1:T, -i})  &=& P({\bf y}_{1,i} |  {\bf y}_{1,-i})
 P({\bf y}_{2,i} | {\bf y}_{1}, {\bf y}_{2,-i}) \\
 &&P({\bf y}_{3,i} | {\bf y}_{1:2}, {\bf y}_{3,-i})  \cdots
 P({\bf y}_{T,i} | {\bf y}_{1:T-1}, {\bf y}_{T,-i}). \notag
\end{eqnarray}
Therefore the log-likelihood of observations for tract $i$ conditional on the other observations in cluster $k$ is 
\begin{eqnarray}
\log P({\bf y}_{1:T, i} | {\bf y}_{1:T, -i}) = \sum_{t=1}^T \log P({\bf y}_{t,i} | {\bf y}_{1:t-1}, {\bf y}_{t,-i}).
\end{eqnarray} 
At time $t$, we have the joint distribution ${\bf y}_{t,i}, {\bf y}_{t,-i}| {\bf y}_{1:t-1}$, which is ${\bf y}_t|{\bf y}_{1:t-1}$ in Eq. (\ref{eqnApp: jointYtLikelihood}). We can then derive the conditional distribution ${\bf y}_{t,i} | {\bf y}_{1:t-1}, {\bf y}_{t,-i}$ by the conventional conditional multivariate normal distribution as follows:
\begin{eqnarray}
A|B \sim \mathcal{N} \left(  \mu_A+\Sigma_{AB} \Sigma_{BB}^{-1}(B-\mu_B),
\Sigma_{AA} - \Sigma_{AB} \Sigma_{BB}^{-1} \Sigma_{BA} \right)
\end{eqnarray}
for the general form of a joint multivariate normal distribution  
\begin{eqnarray}
\left(
\begin{array}{c}
A\\
B
\end{array}
\right) \sim \mathcal{N} 
\left[
\left(
\begin{array}{c}
\mu_A\\
\mu_B
\end{array}
\right),
\left(
\begin{array}{cc}
\Sigma_{AA}, \Sigma_{AB}\\
\Sigma_{BA}, \Sigma_{BB}
\end{array}
\right)
\right].
\end{eqnarray}

\subsection{Sufficient statistic Kalman filter}\label{AppSub:condLike_suffStat}
If all $p^{(k)}$ tracts in a particular cluster $k$ have observations at time $t$, the sufficient statistic multivariate Kalman filter algorithm is as follows:
\begin{align}
	\begin{aligned}
\textrm{Predict} \quad \mu_{t|t-1} & =  A \mu_{t-1|t-1}\\
V_{t|t-1} & =  AV_{t-1|t-1}A^T + Q\\
\textrm{Calculate} \quad S_t & =  V_{t|t-1}+ \bar{R}_t\\
\textrm{(Kalman gain matrix)} \quad K_t &= V_{t|t-1} S_t^{-1}\\
\textrm{Filter} \quad \mu_{t|t} &= \mu_{t|t-1} + K_t (\bar{\dot{\bf y}}_t - \mu_{t|t-1})\\
V_{t|t} &= \left( {\bf I} - K_t\right) V_{t|t-1} 
\end{aligned}
\end{align}
where $\dot{\bf y}_t$ denotes the vector of observations with hedonic effects removed and $\bar{\dot{\bf y}}_t$ the tract-specific mean of all observations at time $t$ after removing hedonic effects. The matrix $\bar{R}_t$ is the diagonal matrix of size $p^{(k)}$-by-$p^{(k)}$ with $(i,i)$-th entry being $\sigma_i^2/L_{t,i}$. The variable $\sigma_i^2$ is the observational variance for tract $i$ and the variable $L_{t,i}$ is the number of observations in tract $i$ at time $t$. Note that all matrix operations above are of the size of the cluster, $p^{(k)}$. If some tracts at time $t$ have no transactions, i.e. $L_{t,i}=0$, we use the following recursion instead: 
\begin{align}
	\begin{aligned}
\textrm{Predict} \quad \mu_{t|t-1} & =  A \mu_{t-1|t-1}\\
V_{t|t-1} & =  AV_{t-1|t-1}A^T + Q\\
\textrm{Calculate} \quad S_t & =  \bar{C}_t V_{t|t-1} \bar{C}_t^T + \bar{R}_t\\
\textrm{(Kalman gain matrix)} \quad K_t &= V_{t|t-1} \bar{C}_t^T S_t^{-1}\\
\textrm{Filter} \quad \mu_{t|t} && \mu_{t|t-1} + K_t (\bar{\dot{\bf y}}_t - \mu_{t|t-1})\\
V_{t|t} &= \left( {\bf I} - K_t \bar{C}_t \right) V_{t|t-1} 
\end{aligned}
\end{align}
In the formula above, $\bar{C}_t$ is an indicator matrix of non-zero sales at time $t$. The matrix has size $p^{(k)} \prime \times p^{(k)}$, where $p^{(k)} \prime$ is the number of tracts that have observations at time $t$ (therefore $p^{(k)} \prime \leq p^{(k)}$). The response variance matrix $\bar{R}_t$ is of size $p^{(k)} \prime \times p^{(k)} \prime$ and includes the variance for tracts that have observations.\\
\\
\section{Derivation of Sampling Steps}
In this section, we provide further details and derivations of the sampling steps outlined in Section~\ref{sec6:posterior} of the main paper.

\subsection{Forward filter backward sampler for the intrinsic price dynamics}\label{App:sampleX}
To sample the latent state sequence, we run a forward filter backward sampler.\\
{\bf Forward Kalman Filter}
\begin{enumerate}
\item Initialize filter with $\mu_{0|0}, V_{0|0}$, where $ X_0 \sim N(\mu_{0|0}, V_{0|0}) $
\item Working forward in time, for $t=1, \cdots, T$, implement the sufficient statistic filter of Appendix \ref{AppSub:condLike_suffStat} to obtain $\mu_{t|t}, V_{t|t}$ for $t=1, \cdots, T$, where $X_t|{\bf y}_{1:t} \sim N(\mu_{t|t}, V_{t|t})$.
\end{enumerate}

{\bf Backward Sampler}{
\begin{enumerate}
\item
Draw $X_T$ from $P(X_T|{\bf y}_{1:T})=N(\mu_{T|T}, V_{T|T})$.
\item
Sequentially sample backward, for $t=T-1, \cdots, 0$,
$x_t$ from $P(X_t| x_{t+1}, {\bf y}_{1:t})$:
\begin{eqnarray}
x_t \sim N\left[\mu_{t|t} + J_t (x_{t+1} - \mu_{t+1|t}), \quad V_{t|t} - J_t V_{t+1|t} J_t^T \right]
\end{eqnarray}
where $J_t = V_{t|t} A^T V_{t+1|t}^{-1}$.\\
\end{enumerate}
}
\subsection{Sampling the latent factor ${\boldsymbol \eta^*}$}\label{App:sampleEta}
For any $t$, the vector of latent states for all $p$ tracts jointly follows a vector autoregressive (VAR) process as follows:
{\small
\begin{eqnarray}
\left[\begin{array}{l}
x_{t,1}\\ \vdots \\ x_{t,n}
\end{array}\right]
& =& \left(\begin{array}{lll}
 a_1 &  & 0 \\
  & \ddots& \\
 0 &  & a_n
 \end{array} \right)
 \left[\begin{array}{l}
 x_{t-1,1}\\ \vdots \\ x_{t-1,n}
 \end{array}\right]
  + (\Lambda \cdot Z ) 
 \left[\begin{array}{l}
 \eta^*_{t,1}\\
 \vdots\\
 \eta^*_{t,K}
 \end{array}\right]
  + \tilde{\boldsymbol \epsilon}_t.
 \end{eqnarray}}
The VAR process can be written in the form of vectors and matrices:
 \begin{eqnarray}
{\bf x_t} = A {\bf x_{t-1}} + (\Lambda \cdot Z) {\boldsymbol \eta^*}_t + \tilde{\boldsymbol \epsilon}_t.
\end{eqnarray}
such that
\begin{eqnarray}
{\bf x_t}-A{\bf x_{t-1}} \sim \mathcal{N}_n \left[  (\Lambda \cdot Z) {\boldsymbol \eta^*}_t, \; \sigma_0^2 I_n\right].
\end{eqnarray}
By first multiplying $(\Lambda \cdot Z)^T$, we get
{\small
\begin{eqnarray}
\qquad (\Lambda \cdot Z)^T ({\bf x_t}-A{\bf x_{t-1}}) &\sim& \mathcal{N}_K \left[ (\Lambda \cdot Z)^T (\Lambda \cdot Z) {\boldsymbol \eta^*}_t, \; \sigma_0^2 (\Lambda \cdot Z)^T(\Lambda \cdot Z)\right].
\end{eqnarray}}
We then multiply by $\left[(\Lambda \cdot Z)^T(\Lambda \cdot Z)\right]^{-1}$ and obtain
{\footnotesize
\begin{eqnarray}
\qquad \left[(\Lambda \cdot Z)^T(\Lambda \cdot Z)\right]^{-1}  (\Lambda \cdot Z)^T ({\bf x_t}-A{\bf x_{t-1}}) 
\sim \mathcal{N}_K \left\{ {\boldsymbol \eta^*}_t,  \; \sigma_0^2 \left[(\Lambda \cdot Z)^T(\Lambda \cdot Z)\right]^{-1}\right\} \label{eqnApp:likelihoodEta}
\end{eqnarray}}
Given the prior of ${\boldsymbol \eta^*}_t \sim \mathcal{N}_K \left( {\bf 0}, I_n \right) $ and the likelihood in Eq. (\ref{eqnApp:likelihoodEta}) ,
by conjugacy, the full conditional distribution for ${\boldsymbol \eta^*}_t$ is
{\small
\begin{eqnarray}
{\boldsymbol \eta^*}_t \left| {\boldsymbol \lambda},{\bf z,x}, \sigma_0^2 \right. 
\sim  \mathcal{N}_K
\left\{ 
\begin{array}{cc}
V\frac{1}{\sigma_0^2}(\Lambda \cdot Z)^T ({\bf x_t}-A{\bf x_{t-1}}) ,\\
V = \left[ I_K + \frac{1}{\sigma_0^2}(\Lambda \cdot Z)^T(\Lambda \cdot Z) \right]^{-1}
\end{array}
\right\}
\end{eqnarray}}

%
%
\subsection{Sampling the factor loadings $\lambda$} \label{App:sampleLambda}
For any $t$, 
\begin{eqnarray}
x_{t,i} = a_i x_{t-1, i} + \lambda_{ik}Z_{ik}\eta^*_{t,k} + \tilde{\epsilon}_{t,i}
\end{eqnarray}
If $Z_{ik}=0$, then the full conditional distribution for $\lambda_{ik}$ is just its prior,
\begin{eqnarray}
\lambda_{ik}|{\bf x, a, \boldsymbol{\eta^*}, z}, \sigma_0^2 \sim \mathcal{N}(\mu_{\lambda}, \sigma^2_{\lambda})
\end{eqnarray} 
If $Z_{ik} =1$ then
{\small 
\begin{eqnarray}
&& p(\lambda_{ik}|{\bf x, a, \boldsymbol{\eta^*}, z}, \sigma_0^2)  \\
&\propto& \mathcal{N} (\mu_{\lambda}, \sigma^2_{\lambda}) 
\prod_{t=1}^T \mathcal{N} (x_{t,i}|a_ix_{t-1,i}+\lambda_{ik}\eta^*_{t,k}, \sigma_0^2) \nonumber \\
& \propto & \mathcal{N} (\mu_{\lambda}, \sigma^2_{\lambda}) 
\prod_{t=1}^T \mathcal{N} \left( \left. \frac{x_{t,i}-a_ix_{t-1,i}}{\eta^*_{t,k}} \right|\lambda_{ik} , \frac{\sigma_0^2}{\eta_{t,k}^{*2}}\right)  \nonumber\\
&\propto& \mathcal{N} \left[ v\left( \frac{\mu_\lambda}{\sigma_\lambda^2}+\sum_{t=1}^T \frac{(x_{t,i}-a_ix_{t-1,i})/\eta^*_{t,k}}{\sigma_0^2/\eta_{t,k}^{*2}} \right),
v=\left( \frac{1}{\sigma_\lambda^2}+\frac{1}{\sigma_0^2} \sum_{t=1}^T \eta_{t,k}^{*2} \right)^{-1} \right] \nonumber
\end{eqnarray}}
In summary, the full conditional distribution for $\lambda_{ik}$ is
\begin{eqnarray}
\lambda_{ik}|{\bf x, a, \boldsymbol{\eta^*}, z}, \sigma_0^2 \sim \mathcal{N}(\mu_{ik}^*, v_{ik}^*)
\end{eqnarray}
where 
{\small
\begin{eqnarray}
(\mu_{ik}^*, v_{ik}^*)=\left\{ \begin{array}{ll}
\mu_\lambda, \sigma_\lambda^2 & \textrm{if } Z_{ik}=0\\
v\left(\mu_\lambda \frac{1}{\sigma_\lambda^2}+\frac{1}{\sigma_0^2} \sum_{t=1}^T \epsilon_{t,i}\eta^*_{t,k}\right),
v=\left( \frac{1}{\sigma_\lambda^2}+\frac{1}{\sigma_0^2} \sum_{t=1}^T \eta_{t,k}^{*2} \right)^{-1} & \textrm{if } Z_{ik}=1
\end{array} \right. .  \notag
\end{eqnarray}}
Here $\epsilon_{t,i}=x_{t,i}-a_ix_{t-1,i}$ and $\sum_{t=1}^T \epsilon_{t,i}\eta^*_{t,k}$ can be written as the inner product ${\boldsymbol{\epsilon}_i^T \boldsymbol{\eta^*}_k}$.

\subsection{Sampling the autoregressive process parameters $a_i$}\label{App:SampleE}
By the likelihood in Eq. (\ref{eqn:simpleMod1}) and Eq. (\ref{eqn:simpleMod3}) of the main paper,  for $z_i=k$
\begin{eqnarray}
x_{t,i} = a_i x_{t-1,i} + \lambda_{ik} \eta^*_{t,k} + \tilde{\epsilon}_{t,i}, \quad \tilde{\epsilon}_{t,i} \sim \mathcal{N}(0, \sigma_0^2).
\end{eqnarray}
Therefore,
\begin{eqnarray}
x_{t,i} \sim \mathcal{N}(a_i x_{t-1,i} + \lambda_{ik} \eta^*_{t,k}, \; \sigma_0^2).
\end{eqnarray}
By rearranging the terms, we get
\begin{eqnarray}
\frac{x_{t,i}-\lambda_{ik} \eta^*_{t,k}}{x_{t-1,i}} \sim \mathcal{N}\left(a_i, \; \frac{\sigma_0^2}{x_{t-1,i}^2}\right), \quad \textrm{i.i.d. for } t=1,\cdots, T.
\end{eqnarray}
By conjugacy, the posterior distribution of the AR process coefficient $a_i$ is
\begin{align}
p(a_i \left|\mu_a, \sigma_a^2, {\mathbf x}_i, \boldsymbol{\eta^*}^{(k)}, \lambda_{ik},\sigma_0^2, z_i \right.)
&\propto \mathcal{N} (a_i|\mu_a, \sigma_a^2) \prod_{t=1}^T 
\mathcal{N} \left[ \frac{x_{t,i}-\lambda_{ik} \eta^*_{t,k}}{x_{t-1,i}} \left| a_i, \frac{\sigma_0^2}{x_{t-1,i}^2}\right. \right] \\
&\hspace{-2in}\propto \mathcal{N}\left( V\left[\frac{1}{\sigma_a^2} \cdot \mu_a 
+ \sum_{t=1}^T \left(\frac{x_{t-1,i}^2}{\sigma_0^2} \cdot \frac{x_{t,i}-\lambda_{ik}\eta^*_{t,k}}{x_{t-1,i}} \right) \right], 
V=\left( \frac{1}{\sigma_a^2}+\sum_{t=1}^T \frac{x_{t-1,i}^2}{\sigma_0^2} \right)^{-1} \right). \notag
\end{align}

%
%
\subsection{Sampling the covariate parameters $\beta_{i,h}$} \label{App:SampleF}
For tract $i$ and hedonic covariate $h$, the posterior distribution for covariate effect $\beta_{i,h}$ is
\begin{eqnarray}
&&p(\beta_{i,h}|\mu_h, \sigma_h^2, R_i, {\bf x}_{1:T, i}, {\bf y}_{1:T,i}) \nonumber \\
&\propto& N(\beta_{i,h}|\mu_h, \sigma_h^2)
\prod_{t=1}^T\prod_{l=1}^{L_t}N\left(y_{t,i,l}\left|x_{t,i}+\sum_{s \neq h} \beta_s U_{l,s}+\beta_h U_{l,h}\right., R_i\right)  \nonumber\\
&\propto& N(\beta_{i,h}|\mu_h, \sigma_h^2)
\prod_{t=1}^T\prod_{l=1}^{L_t}N\left(\left.y_{t,i,l} - x_{t,i}+\sum_{s \neq h} \beta_s U_{l,s}\right|\beta_h U_{l,h}, R_i\right)  \nonumber \\
&\propto& N(\beta_{i,h}|\mu_h, \sigma_h^2)
\prod_{t=1}^T\prod_{l=1}^{L_t}N\left[\frac{1}{U_{l,h}}\left.\left(y_{t,i,l} - x_{t,i}+\sum_{s \neq h} \beta_s U_{l,s}\right) \right|\beta_h, \frac{R_i}{U_{l,h}^2}\right] \nonumber \\
&\propto& N\left\{ 
\begin{array}{cc}
v\left[  \frac{1}{\sigma_h^2} \mu_h+ \frac{1}{R_i} \sum_{t=1}^T\sum_{l=1}^{L_t}U_{l,h} \left(y_{t,i,l}-x_{t,i}-\sum_{s\neq h}\beta_s U_{l,s}\right) \right] , \\
v = \left(\frac{1}{\sigma_h^2} + \frac{1}{R_i}\sum_{t=1}^T\sum_{l=1}^{L_t}U_{l,h}^2\right)^{-1}
\end{array}
\right\} \notag
\end{eqnarray}
%
%
\subsection{Sampling the DP hyperparameter $\alpha$}\label{App:SampleG}

Following \cite{Escobar94bayesiandensity} and \cite{CrimePaper2013}, we assume a gamma distribution prior for the concentration parameter $\alpha \sim \textrm{Gamma}(\alpha_\alpha, \beta_\alpha)$. We sample an auxiliary variable $\kappa$ to help us sample $\alpha$:
\begin{enumerate}
\item Sample $\kappa \sim \textrm{Beta}(\alpha+1, n)$, where $n$ is the total number of tracts.
\item Sample $\alpha$ from the a mixture of two gamma distributions as follows:
\begin{eqnarray}
\alpha | \kappa, K &\sim& \pi \textrm{Gamma}\left(\alpha_\alpha+K, \beta_\alpha - \log(\kappa)\right) \notag \\
&& + (1-\pi) \textrm{Gamma}\left(\alpha_\alpha+K-1, \beta_\alpha - \log(\kappa)\right), \notag
\end{eqnarray}
where $K$ is the number of unique clusters, and the mixture weight $\pi$ is defined by $\pi/(1-\pi) = (\alpha_\alpha+K-1)/\left( p \left[\beta_\alpha - \log(\kappa)\right]\right)$.
\end{enumerate}


\section{Parallel DPMM sampler}\label{App:parallelDPMM}
Sampling the cluster membership $z_i$ in parallel includes the following two steps:
\begin{enumerate}
\item \textbf{Local step} on each machine in parallel:\\
Conditioned on the processor assignments $\boldsymbol{\gamma}$, we sample the cluster assignments $\{z_i: \gamma_i=j\}$ as in a conventional Dirichlet process mixture model (Section \ref{sec6.1:sampleZ} of the main paper) with concentration parameter $\alpha/P$, for data points assigned to a machine $j$. Since the DPMMs are independent given the processor allocations, we can sample $\{z_i: \gamma_i=j\}$ in parallel across machines.

\item \textbf{Global step} over machines:\\
Each cluster is associated with a single processor. One processor can have multiple clusters. We jointly resample the processor allocations of all data points within a given cluster. We use a Metropolis-Hastings step with a proposal distribution that independently assigns cluster $k$ to processor $j$ with probability $1/P$. This means our accept/reject ratio depends only on the ratio of the likelihoods of the current processor assignments $\{\gamma_i\}$ and the proposal $\{\gamma_i^*\}$.

The likelihood ratio is given by:
\begin{eqnarray}
\frac{p\left(\{\gamma_i^*\}\right)}{p\left(\{\gamma_i\}\right)}
&=& \frac{p\left(\{z_i\}|\gamma_i^*\right)p\left(\{\gamma_i^*\}|\alpha, P\right)}{p\left(\{z_i\}|\gamma_i\right)p\left(\{\gamma_i\}|\alpha, P\right)}\\
&=& \prod_{j=1}^P \prod_{i=1}^{\max(N_j, N_j^*)} \frac{a_{ij} !}{a_{ij}^* !}
\end{eqnarray}

where $N_j$ is the number of data points on machine $j$, and $a_{ij}$ is the number of clusters of size $i$ on machine $j$. The derivation is shown in the supplementary material of \cite{WilliamsonParallelMCMC2013}.

\end{enumerate} 

\section{Hyperprior Settings}

\subsection{Hyperprior for $\sigma_0^2$}\label{App:hyperPriorSigma0Squared}
We set the hyper priors for $\sigma_0^2 \sim \textrm{IG}(\alpha_{\epsilon 0} , \beta_{\epsilon 0})$ with hyperparameters $\alpha_{\epsilon 0} = 0.5, \beta_{\epsilon 0} = 1$. When examining the housing data, for numerical stability we multiply the observations $\left[\log(\textrm{Price}_{t,i,l})-\log(g_t) \right]$ by a factor of $200$. As a result, $99\%$ of outcome values are covered by the interval $[-1.10, 1.61]$. The chosen hyper prior has a long and flat tail distribution over the range of variance. 

\subsection{Hyperprior for $R_{i}$}\label{App:hyperPriorAlphaBetaR0}
We set the hyper priors for $R_i \sim \textrm{IG}(\alpha_{R 0} , \beta_{R 0})$ with hyperparameters $\alpha_{R 0} = 3, \beta_{R 0} = 1$. The chosen hyper prior has a long and flat tail distribution over the range of variance.

%
%
\section{Extended Simulation Results}\label{App:SimulationResults}
\begin{figure}[t!]
\centering
\includegraphics[width=0.9\columnwidth]{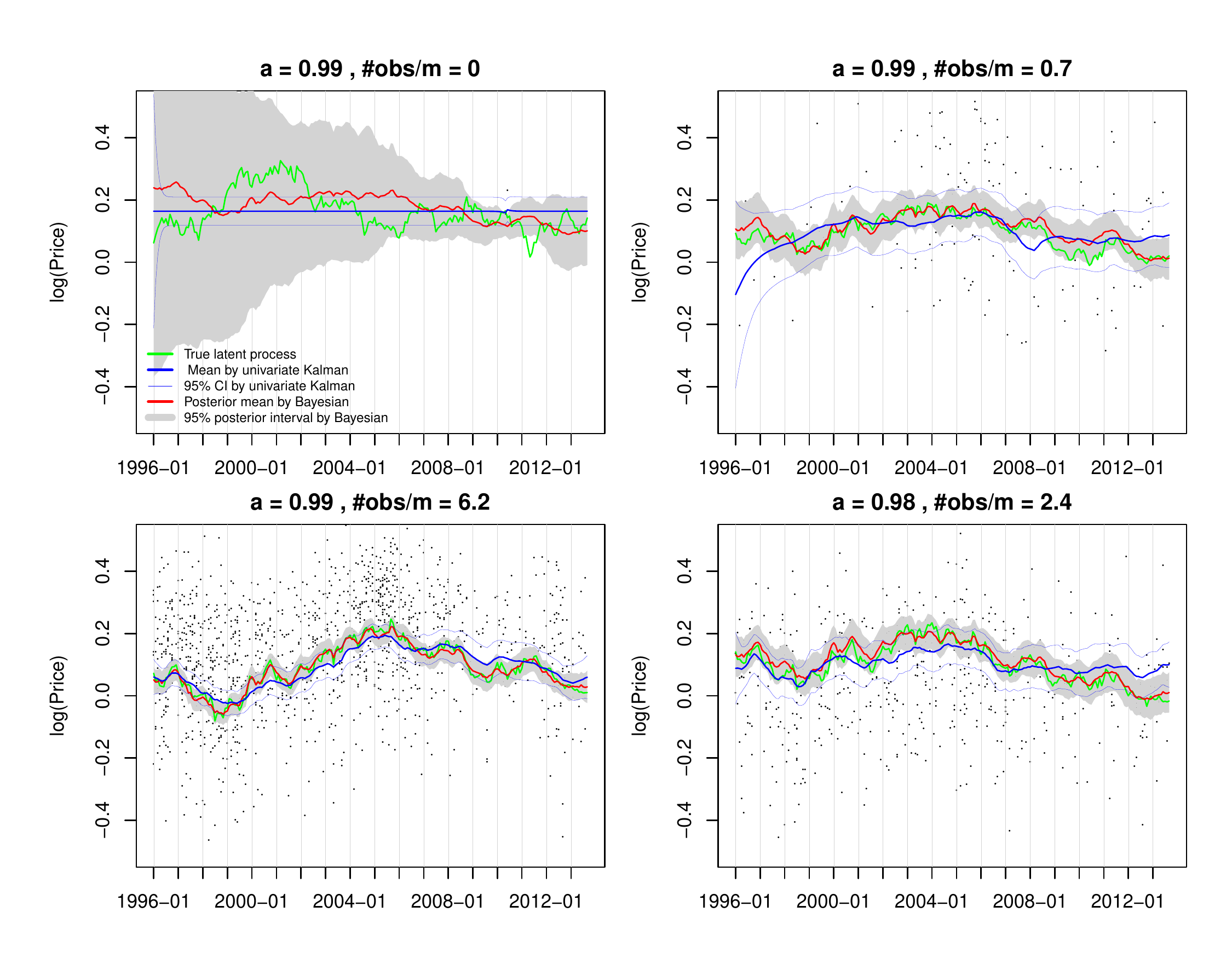}
\caption{Performance of estimating the latent process for Cluster 2.}
\label{fig:recoverTrueX2}
\end{figure}

\begin{figure}[t!]
\centering
\includegraphics[width=0.9\columnwidth]{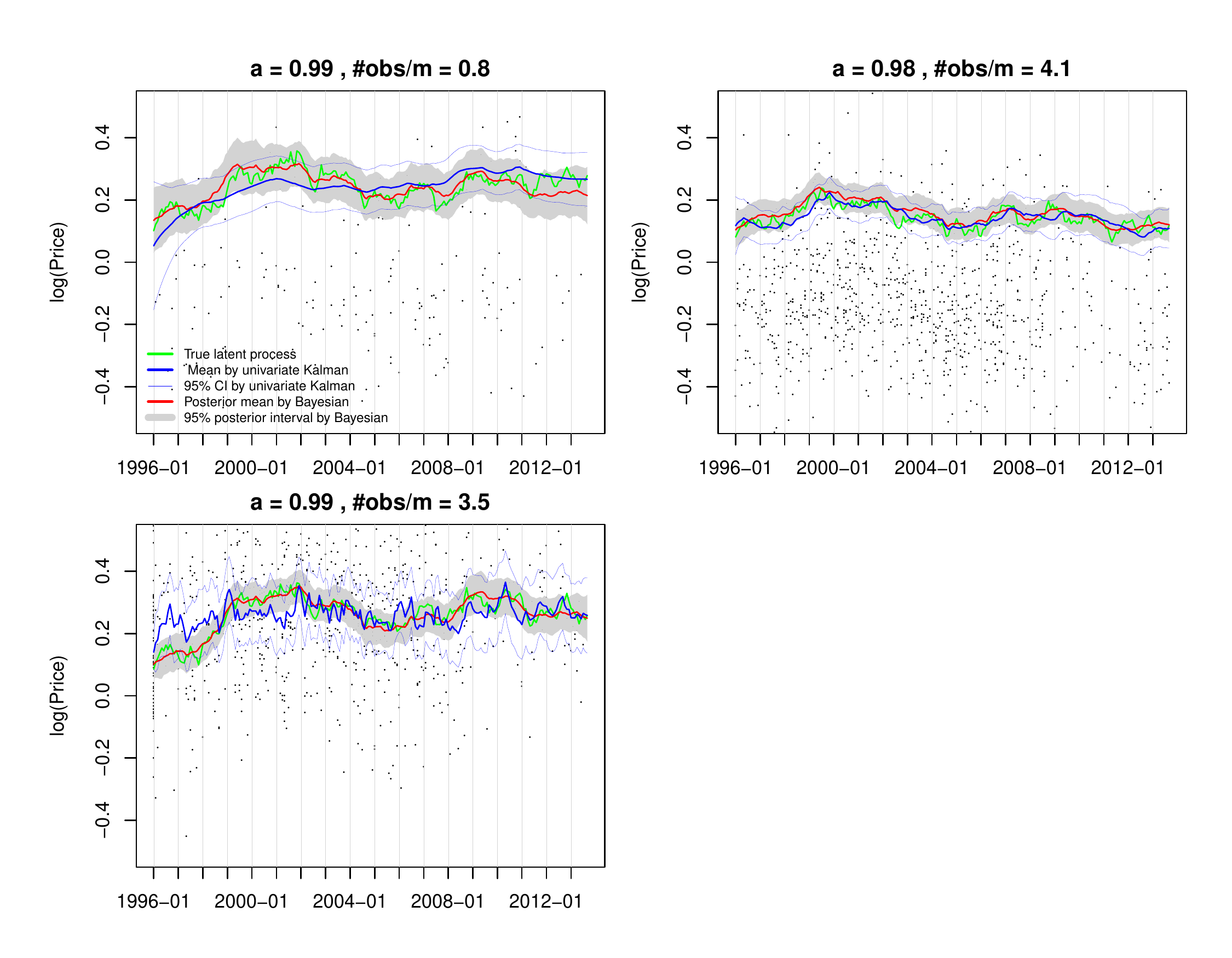}
\caption{Performance of estimating the latent process for Cluster 3.}
\label{fig:recoverTrueX3}
\end{figure}

\begin{figure}[t!]
\centering
\includegraphics[width=0.9\columnwidth]{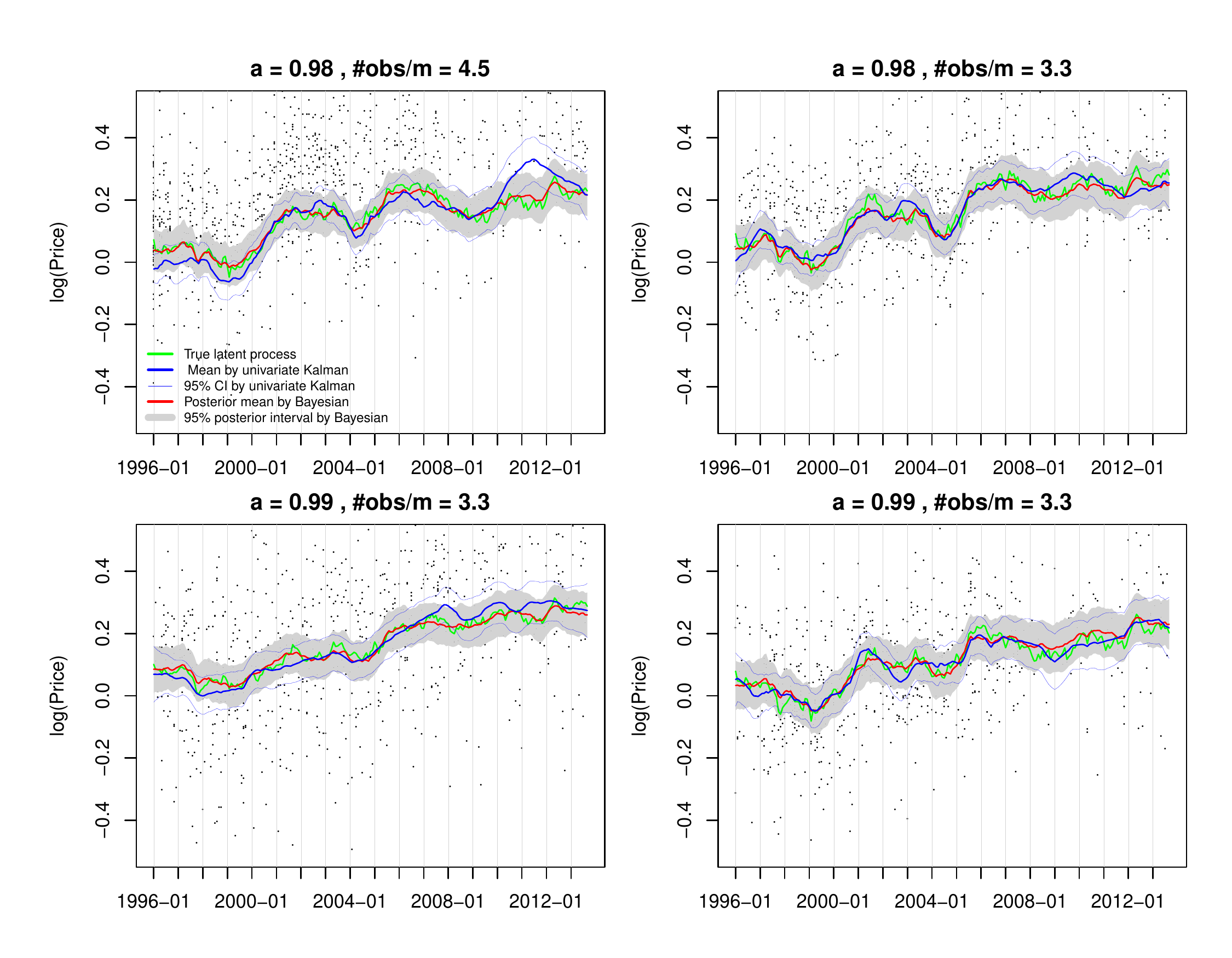}
\caption{Performance of estimating the latent process for Cluster 4.}
\label{fig:recoverTrueX4}
\end{figure}

In this section, we provide a performance analysis of the remaining clusters not examined in Section \ref{sec8.2:simulationResults} of the main paper. Figures \ref{fig:recoverTrueX2}, \ref{fig:recoverTrueX3} and \ref{fig:recoverTrueX4} directly parallel Figure \ref{fig:recoverTrueX} of the main paper and show our performance in estimating the simulated intrinsic price dynamics compared to an independent Kalman-filter-based analysis of the tracts.

%
%
\section{Extended Seattle City Results}\label{App:realDataResults}
\begin{figure}[t!]
\centering
\includegraphics[width=1\linewidth]{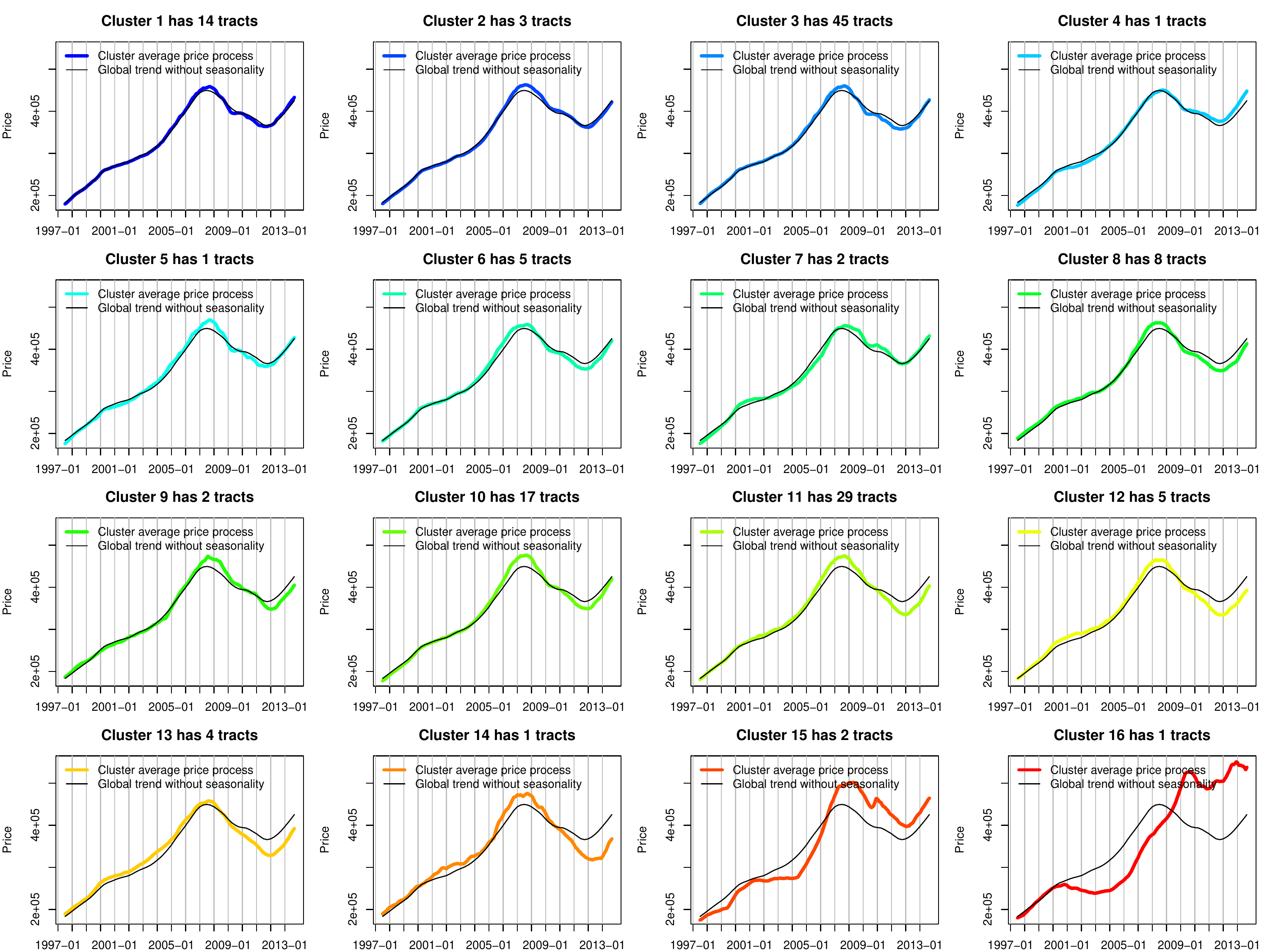}
\caption{Under the MAP sample, we first compute the cluster-average intrinsic price dynamics by averaging $x_{1:T,i}$ over all $i$ with $z_i = k$ for $k = 1,...,16$ (all of the estimated clusters). We then add this cluster-average price to the global trend without seasonality (\emph{various colors}) and hold on the seasonally adjusted global trend (\emph{black}) for comparison.}
\label{fig:ClusterPanel_GlobalTrend_nonSeasonal}
\end{figure}

\begin{figure}[t!]
\centering
\includegraphics[width=0.7\linewidth]{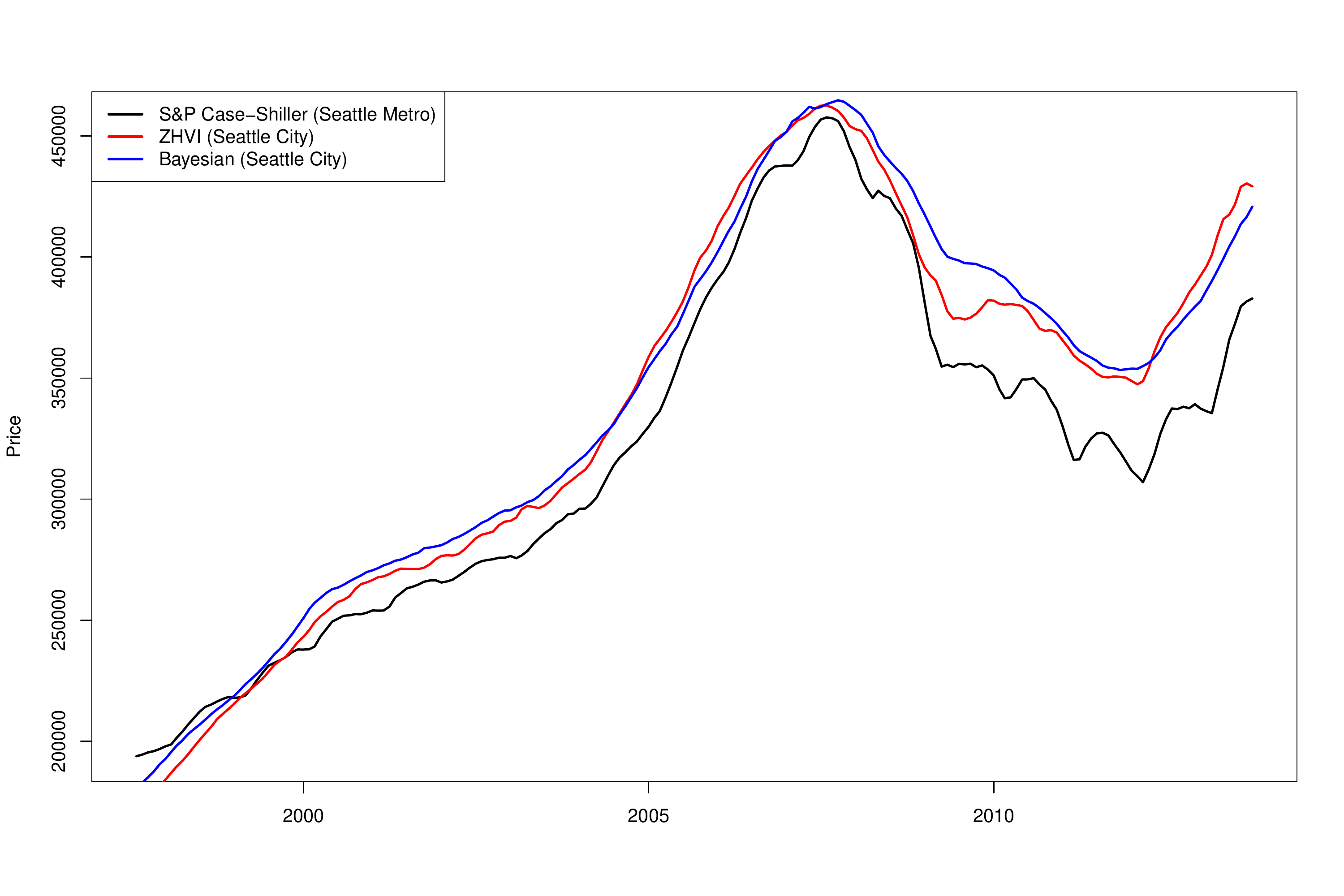}
\caption{Seattle City Price Index by S\&P Case--Shiller, Zillow Home Value Index (ZHVI) and our proposed Bayesian method.}
\label{fig:indexComparison}
\end{figure}

In this section, we present a set of figures from our Seattle City data analysis to augment those presented in the main paper. For the MAP MCMC sample, Figure \ref{fig:ClusterPanel_GlobalTrend_nonSeasonal} displays the average of the intrinsic price processes within a cluster, for each of the 16 inferred clusters.  This plot parallels that of Figure~\ref{fig:ClusterMap} of the main paper, but here in the raw price space instead of $\log$ space and with the global trend added without the seasonality component (for clarity).  We additionally hold on the estimated global trend without seasonality for comparison.  In Figure \ref{fig:indexComparison}, we compare the resulting housing index produced by S\&P Case--Shiller, Zillow Home Value Index (ZHVI), and our Bayesian method at the Seattle City level.  

\subsection{Sales volume and variance over time}\label{App:salesVolume}
\begin{figure}[t!]
\centering
\includegraphics[width=0.9\linewidth]{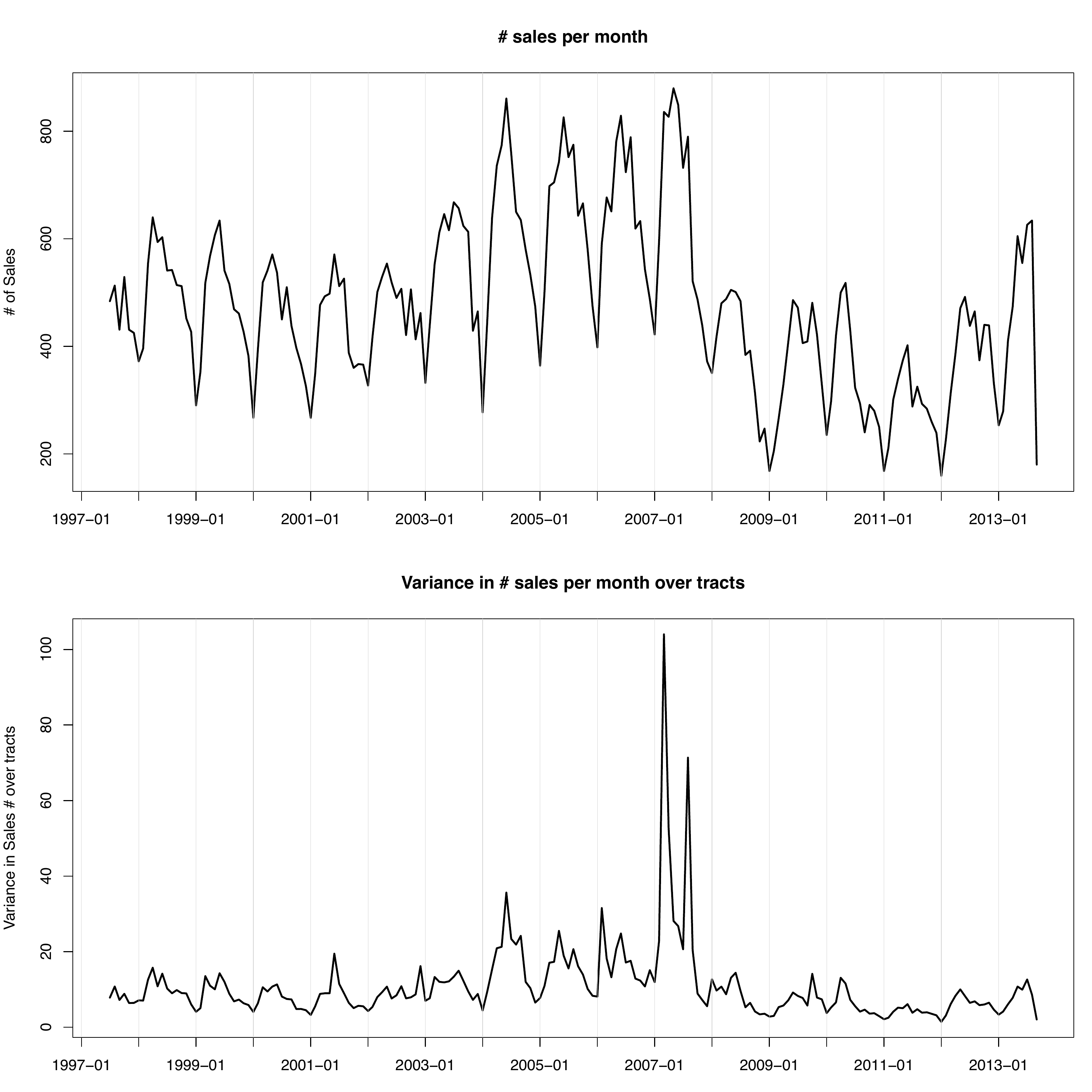}
\caption{Sales volume (\emph{top}) and variance (\emph{bottom}) versus time.}
\label{fig:compareIndex_salesVolume}
\end{figure}

Figure \ref{fig:compareIndex_salesVolume} shows the sales volume and its variance over time, as discussed in Section \ref{sec9.1:SeattleResultsComparison}, together with Figure~\ref{fig:compareZipIndex_threeMethods}, of the main paper. The market boom, roughly 2006-2007, and subsequent stabilization, roughly 2010-2011, were manifested in the different housing sectors in disparate ways. The index formed from the model based on DP clustering is able to capture the dynamics of the change in value for different housing sectors during these two periods.


\end{document}